\documentclass[a4paper,11pt]{article}
\pdfoutput=1
\usepackage{jheppub}

\usepackage{amssymb,amsmath,amsfonts}
\usepackage[normalem]{ulem}
\usepackage[utf8x]{inputenc}
\usepackage{slashed}
\usepackage{graphicx}
\usepackage{tabularx}
\usepackage{here}
\usepackage{color}
\usepackage{csquotes} %fix backwards quotes
\usepackage{comment}
\usepackage{mathrsfs}
\usepackage{float}
\usepackage{ascmac}
\usepackage{multirow}
\usepackage{longtable}
\usepackage{bm}
\usepackage{ulem}

\usepackage[italicdiff]{physics}
\usepackage[hang,small,bf]{caption}
\usepackage[subrefformat=parens]{subcaption}
\makeatletter
\newcommand*\rel@kern[1]{\kern#1\dimexpr\macc@kerna}
\newcommand*\widebar[1]{%
  \begingroup
  \def\mathaccent##1##2{%
    \rel@kern{0.8}%
    \overline{\rel@kern{-0.8}\macc@nucleus\rel@kern{0.2}}%
    \rel@kern{-0.2}%
  }%
  \macc@depth\@ne
  \let\math@bgroup\@empty \let\math@egroup\macc@set@skewchar
  \mathsurround\z@ \frozen@everymath{\mathgroup\macc@group\relax}%
  \macc@set@skewchar\relax
  \let\mathaccentV\macc@nested@a
  \macc@nested@a\relax111{#1}%
  \endgroup
}
\makeatother

\numberwithin{equation}{section}

\preprint{
\begin{minipage}{5cm}
\small
\flushright
EPHOU-25-003\\
KYUSHU-HET-312
\end{minipage}}

\title{More about quark Yukawa textures from selection rules without group actions}

\author{Tatsuo Kobayashi$^{1}$,} 
\author{Yume Nishioka$^{1}$,}
\author{Hajime Otsuka$^{2}$, and} 
\author{Morimitsu Tanimoto$^{3}$} 
\affiliation{
$^1$Department of Physics, Hokkaido University, Sapporo 060-0810, Japan}
\affiliation{
$^2$Department of Physics, Kyushu University, 744 Motooka, Nishi-ku, Fukuoka 819-0395, Japan}
\affiliation{$^3$Department of Physics, Niigata University, Ikarashi 2-8050, Niigata 950-2181, Japan}
\emailAdd{kobayashi@particle.sci.hokudai.ac.jp}
\emailAdd{y-nishioka@particle.sci.hokudai.ac.jp}
\emailAdd{otsuka.hajime@phys.kyushu-u.ac.jp}
\emailAdd{morimitsutanimoto@yahoo.co.jp}

\abstract{
We study the coupling selection rules associated with non-group symmetries, i.e., $\mathbb{Z}_2$ gauging of $\mathbb{Z}_M$ symmetries.
We clarify which Yukawa textures can be derived by our selection rules for $M=3, 4$, and 5, and obtain various textures including the nearest neighbor interaction type and its extension. Some of them cannot be realized by a conventional group-like symmetry. 
They lead to interesting phenomenology such as a solution to the strong CP problem without axion.
}

\makeatletter
\gdef\@fpheader{}
\makeatother

\begin{document}

\maketitle

%%%%%%%%%
\section{Introduction}

Symmetries are important in physics.
In particle physics, symmetries determine selection rules of allowed couplings and forbidden couplings.
We often consider group symmetries in theory leading to definite coupling selection rules.
In models with group symmetry, each field corresponds to a representation of the group.
Their couplings are allowed only if a product of their representations becomes the identity of the group.
When we choose representations of $(n-1)$ fields in the $n$-point coupling, the representation, which the $n$-th field should have in the allowed coupling, is fixed automatically.
When we apply such a group selection rule to the Yukawa couplings, the Yukawa matrices must be diagonal for flavors with different quantum numbers of the group, and we can not realize flavor mixing.
In order to realize the flavor mixing, 
the flavor group must be broken, e.g. by flavon fields.
Effective Yukawa couplings including vacuum expectation values (VEVs) of 
flavons behave as spurions with non-trivial representations of the group.
Note that flavor mixing appears after group symmetry breaking in Abelian flavor group models \cite{Froggatt:1978nt} and non-Abelian ones \cite{Altarelli:2010gt,Ishimori:2010au,Hernandez:2012ra,King:2013eh,King:2014nza,Petcov:2017ggy,Kobayashi:2022moq}.
One exception is modular flavor symmetric models \cite{Feruglio:2017spp}.
In those models, Yukawa couplings are modular forms, which are holomorphic functions of modulus, and transform non-trivially 
under finite modular groups such as $S_3$, $A_4$, $S_4$, and $A_5$ \cite{Feruglio:2017spp,Kobayashi:2018vbk,Penedo:2018nmg,Novichkov:2018nkm}. (See for reviews Refs.~\cite{Kobayashi:2023zzc,Ding:2023htn}, including a possible set of ultraviolet completions.)

Recently, the symmetries, which have no group structure, were studied intensively.
They satisfy the fusion algebra $U_i \otimes U_j=\sum_k c^k_{ij} U_k$, where $U_i$ are topological operators.
These symmetries were discussed in two-dimensional conformal field theories and four-dimensional quantum field theories. 
More than one operator appears on the right-hand side of the above fusion algebra.
On the other hand, in group theory, the product of two group elements $a$ and $b$ is uniquely fixed, i.e., $ab=c$.
That is an important difference.
Thus, it would be possible to derive flavor mixing from symmetry without group structure, even if such a symmetry is not broken. 
Such a situation happens when the four-dimensional theory has, for instance, a non-invertible symmetry, 
where fields $\phi_i$ can be labeled by conjugacy classes $[g_i]$ of a certain group $G$.
Then, one can arrive at the following fusion of two conjugacy classes $[g_i]\otimes [g_j] = \sum_k d_{ij}^k [g_k]$ \cite{Hamidi:1986vh,Dijkgraaf:1989hb}, 
which put constraints on the interaction of fields \cite{McNamara:2021cuo,Thorngren:2021yso,Heckman:2024obe,Kaidi:2024wio}.

In Ref.~\cite{Kobayashi:2024cvp}, we applied this concept to flavor physics.
We studied the symmetries and coupling selection rules, which are obtained by $\mathbb{Z}_2$ gauging of $\mathbb{Z}_M$ symmetries, where $\mathbb{Z}_2$ is the outer automorphism of $\mathbb{Z}_M$ symmetries.
These coupling selection rules can be derived from a certain string compactification as non-invertible symmetries \cite{Kobayashi:2024yqq,Funakoshi:2024uvy}. 
For instance, in the context of four-dimensional (4D) effective action of higher-dimensional Yang-Mills theory compactified on $T^2/\mathbb{Z}_2$ with magnetic fluxes, momentum operators on $T^2$ are constrained in the $\mathbb{Z}_2$-invariant form. 
These topological operators induce the non-invertible symmetries. Then, it was found that the non-invertible symmetries provide flavor symmetries of chiral zero modes and determine their flavor structure. 
The obtained selection rules do not correspond to those associated with group-like symmetries. 
Furthermore, one can derive the flavor mixing without symmetry breaking.
In our approach, various texture zeros can be realized in the $3\times 3$ Yukawa matrices, which can not be obtained from group selection rules before symmetry breaking.
Some of them have not been studied until now  because 
they were not derived from conventional flavor symmetries.

The texture zeros approach has a long history, but it does not have a strong theoretical basis. 
In the framework of two families of quarks, Weinberg put a mass matrix
by hand for the down-type quark 
sector with zero  (1,1) entry in the basis in which the 
up-type quark mass matrix is diagonal \cite{Weinberg:1977hb}.  
%%%%%%%%%%%%%%%%%%%%%%%%%%%%%%
Fritzsch extended the above approach to the three
family case \cite{Fritzsch:1977vd,Fritzsch:1979zq}. 
Ramond, Roberts and Ross have  studied  four or five zeros textures,
which were also put by hand,
for symmetric or hermitian quark mass matrices \cite{Ramond:1993kv}.
Their textures are not viable today since they cannot describe 
the current rather precise data on the CKM quark mixing 
matrix. However, the texture zero approach to the  mass matrices 
of quarks and leptons
is still promising because it has  the big prediction power in the flavor physics \cite{Fritzsch:2002ga,Xing:2015sva,Frampton:2002yf,
Kageyama:2002zw}.

In this paper, we systematically study the texture structures, which can be derived by our coupling selection rules associated with $\mathbb{Z}_2$ gauging of $\mathbb{Z}_M$ symmetries.
Also, we discuss the phenomenological implication
of interesting texture zeros we obtain in the quark sector.

This paper is organized as follows. 
In section \ref{sec:fusion-rule}, we explain how to apply non-trivial fusion rules to particle physics. 
In section \ref{sec:selction-rule}, we review our coupling selection rules.
In section \ref{sec:M=3}, we systematically study which Yukawa textures can be derived for $M=3$.
Similarly, we study textures for $M=4$ and 5 in section \ref{sec:M=4-5}.
In section \ref{sec:phenomenology}, we discuss the phenomenological implications of interesting textures in the quark sector.
Section \ref{sec:con} is our conclusion.
In Appendices \ref{app:M=4} and \ref{app:M=5}, we show our detailed results for $M=4$ and 5.
In Appendix \ref{app:Jcp}, we show the relation about the Jarlskog invariant.
\section{Coupling selection rules and fusion rules}
\label{sec:fusion-rule}

Here, we explain how to apply non-trivial fusion rules to particle physics.
We start with the selection rules due to group theory in order to emphasize the difference of selection rules.

Suppose that the elements, $a,b,c$, belong to the group $G$, and they satisfy the following multiplication law:
\begin{align}
\label{eq:group-product}
    ab=c.
\end{align}
The element appearing in the right hand side is unique in group theory.
In $G$-invariant field theory, each field corresponds to a (representation of) element of $G$.
For example, the scalar fields $\phi_a, \phi_b, \phi_c,\phi_d$ correspond to the elements $a,b,c,d$ in $G$.
If $ab=c$, group theory allows the process $\phi_a + \phi_b \to \phi_c$.
If $ab \neq d$, the process $\phi_a + \phi_b \to \phi_d$ is forbidden.
That is the coupling selection rules due to group theory.
When $G$ is Abelian, that is the charge conservation law.
The important point is that the element appearing in the right hand side of Eq.~(\ref{eq:group-product}) is unique.
Of course, when another field $\phi'_c$ has the same charge as $\phi_c$, the process $\phi_a + \phi_b \to \phi'_c$ is allowed.
There is no difference between $\phi_c$ and $\phi'_c$ in the group $G$.

For example, let us assume the flavor group symmetry $G$.
Three generations of quarks and leptons have different charges under $G$.
We assign their charges as well as the charge of the Higgs field such that diagonal entries are allowed in order to derive their masses.
However, in such a flavor symmetric model, we can not realize flavor mixing, because off-diagonal entries are forbidden by uniqueness of the right hand side element $c$ in Eq.~(\ref{eq:group-product}).

Now, we consider the set of elements $U_i$, i.e., $\{ U_i \}$, which may correspond to operators.
Suppose that they satisfy the following multiplication law:
\begin{align}
    U_i U_j = \sum_k c_{ijk} U_k.
\end{align}
That is the so-called fusion rule.
The elements appearing in the right hand side are not unique in non-trivial fusion rules.
Thus, we can not define the inverse of elements unlike in group theory.
These fusion rules may be originated from string compactifications.
Each field $\phi_i$ corresponds to one element $U_i$.
Then, the above rules determine the coupling selection rule among $\phi_i$.
There are the processes $\phi_i + \phi_j \to \phi_k$ when $c_{ijk} \neq 0$.
Of course, another field $\phi'_k$ can also correspond to the same element $U_k$, and there is no difference between these fields from the viewpoint of the fusion rules like the fields with the same charge in group theory.

As an illustration, we show a simple fusion rule, 
\begin{align}
    U_1U_2=U_3+U_5.
\end{align}
There are the fields $\phi_i$ corresponding to $U_i$ for $i=1,\cdots$.
The processes $\phi_1 + \phi_2 \to \phi_3$ and $\phi_1 + \phi_2 \to \phi_5$ can occur, but the process $\phi_1 + \phi_2 \to \phi_4$ is forbidden by the selection rule due to the above fusion rule.
If another field $\phi'_3$ also corresponds to $U_3$ as $\phi_3$, 
the process $\phi_1 + \phi_2 \to \phi'_3$ can also occur.

The important difference from group theory is that the right hand side in multiplication laws is not unique.
That can lead to off-diagonal entries in mass matrices and flavor mixing as shown in the following sections.
If the right hand side in multiplication laws for all the elements $U_i$ is unique, there would be no difference from the selection rules due to group theory.
The definite coupling selection rules imply symmetries in a theory.
Thus, we call the selection rules due to the above fusion rules as symmetries.
In particular, we study flavor symmetries when we apply some fusion rules to flavor physics.

It is possible to extend symmetries by combining two or more symmetries.
Such a procedure is similar to a direct product of groups, e.g. $G_1 \times G_2$.
For example, in the $G_1\times G_2$-invariant theory, fields correspond to 
some elements of both $G_1$ and $G_2$, e.g. $\phi_{a_1,a_2}$, $\phi_{b_1,b_2}$, $\phi_{c_1,c_2}$, where $a_1,b_1,c_1 \in G_1$ and $a_2,b_2,c_2 \in G_2$.
Their coupling selection rules are determined by two independent multiplications:
\begin{align}
    a_1b_1=c_1, \qquad a_2b_2=c_2,
\end{align}
where the elements $a_1,b_1,c_1$ in $G_1$ and the elements $a_2,b_2,c_2$ in $G_2$.
We can write multiplication laws of $G_1\times G_2$ by
\begin{align}
    (a_1,a_2)(b_1,b_2)=(c_1,c_2).
\end{align}
For instance, in $U(1)_1\times U(1)_2$ theory, each field has two independent charges, $q_1$ and $q_2$.
$U(1)_1\times U(1)_2$ symmetry requires conservation of both charges, independently.
Similarly, it is possible to extend the coupling selection rules due to non-trivial fusion rules.
Suppose that we have two independent fusion rules:
\begin{align}
    U_i U_j = \sum_k c_{ijk} U_k, \qquad  U'_{i'} U'_{j'} = \sum_{k'} c'_{i'j'k'} U'_{k'}.
\end{align}
We combine them as 
\begin{align}
    (U_i,U'_{i'})( U_j,U'_{j'}) = \sum_{k,k'} c_{ijk}c'_{i',j',k'} (U_k, U'_{k'}).
\end{align}
Each field $\phi_{i,i'}$ corresponds to an element, $(U_i, U'_{i'})$.
Their coupling selection rules follow the above fusion rules.
Further, we can combine three and more non-trivial rules, and we can combine non-trivial fusion rules with conventional coupling selection rules due to group theory.

One way to derive non-trivial fusion rules is to start with a group symmetry, e.g. $G$ and then to relate elements of the group $G$ by discrete $\mathbb{Z}_M$ symmetry, which may be (outer) automorphism of $G$, i.e., $\mathbb{Z}_M$ gauging.
Here, we use the fusion rules, which are obtained by $\mathbb{Z}_2$ gauging of $\mathbb{Z}_M$ symmetries.

%%%%%%%%%%%%%%%%%%%%%%%%%%%%%%%%%%%%%%%%%%%%%%%%%%%%%%%%%%%%%%%%%
\section{Coupling selection rules due to $\mathbb{Z}_2$ gauging}
\label{sec:selction-rule}
%%%%%%%%%%%%%%%%%%%%%%%%%%%%%%%%%%%%%%%%%%%%%%%%%%%%%%%%%%%%%%%%%

In this section, we briefly review the selection rule of fields 
labeled by the conjugacy class of a group in the context of 4D quantum field theory (QFT). 
Following Ref. \cite{Kobayashi:2024cvp}, we start with the $\mathbb{Z}_M$ symmetry with generators $g$. 
By introducing the conjugacy classes of the $\mathbb{Z}_M$:
\begin{align}
    g^k,
\end{align}
with $k=0,1,...,M-1$ (mod $M$), their selection rules are given by
\begin{align}
    \tilde{g}^{k_1}\tilde{g}^{k_2} = \tilde{g}^{k_1 +k_2},
\label{eq:2_point}
\end{align}
for some $\Tilde{g}^k\in g^k$ and $\tilde{g}^M=\tilde{g}^0=e$ with $e$ being the identity. 
The field labeled by this conjugacy class $g^k$ has the charge $k$ under the $\mathbb{Z}_M$ symmetry. 
The couplings of $n$ kinds of fields $\{\phi_1, ..., \phi_n\}$ labeled by the conjugacy classes $\{ g^{k_1},..., g^{k_n}\}$ in the 4D QFT are allowed when 
\begin{align}
    \tilde{g}^{k_1}\tilde{g}^{k_2} \cdots \tilde{g}^{k_n} = e,
\label{eq:rule_before}
\end{align}
for some $\Tilde{g}^k\in g^k$. 

Let us consider the (outer) automorphism of $\mathbb{Z}_M$\footnote{Since the inner automorphism of $\mathbb{Z}_M$ is trivial, the automorphism and its outer automorphism of $\mathbb{Z}_M$ are identified with each other.}, namely $\mathbb{Z}_2$ with generators $e$ and $r$;
\begin{align}
    eg e^{-1} = g,\qquad
    rg r^{-1} = g^{-1}.
\end{align}
When the $\mathbb{Z}_2$ symmetry associated with this outer automorphism is gauged, one can define the following class:
\begin{align}
    [g^k] = \{ h g^k h^{-1}\,|\, h=e,r\},
\end{align}
where the index $k$ runs from $0$ to $p$ for both $M=2p$ and $M=2p+1$ with $p\in \mathbb{Z}$. 
Since the class $[g^k]$ includes both representations $g^k$ and $g^{-k}\simeq g^{M-k}$,\footnote{Note that $g^{-k}$ is identified with $g^{M-k}$.} the class has the charge $k$ and $M-k$ in the language of the $\mathbb{Z}_M$ symmetry.

The $\mathbb{Z}_2$ gauging of $\mathbb{Z}_M$ symmetries can be realized by string compactifications.
For example, toroidal compactifications $T^{2n}$ with magnetic fluxes lead to $\mathbb{Z}_M$ symmetries \cite{Abe:2009vi,Berasaluce-Gonzalez:2012abm,Marchesano:2013ega}, where each mode $\varphi_j$ transforms 
\begin{align}
    \varphi_j \to g^{k_j} \varphi_j.
\end{align}
That is, the mode $\varphi_j$ has the $\mathbb{Z}_M$ charge $k_j$.
Instead of toroidal compactifications, we consider its orbifold compactifications, $T^{2n}/\mathbb{Z}_2$.
All of the mode $\varphi_j$ are not invariant under $\mathbb{Z}_2$ orbifolding, but $\mathbb{Z}_2$-invariant modes can be written by \cite{Abe:2008fi}
\begin{align}
    \phi_j=\varphi_j+\varphi_{M-j}.
\end{align}
In general, these modes $\phi_j$ have no definite $\mathbb{Z}_M$ charged unlike $\varphi_j$, but they correspond to the class $[g^k]$ \cite{Kobayashi:2024yqq} and they behave to have both charges, 
$k_j$ and $(M-k_j)$.

The selection rule of this class $[g^k]$ is different from Eq. \eqref{eq:rule_before}. 
To see this structure, let us remind the reader of the selection rule \eqref{eq:rule_before}. 
Before $\mathbb{Z}_2$ gauging, we have the rules,
\begin{align}
\begin{array}{cc}
\label{eq:ZM-rule}
g^{k_1}g^{k_2}=g^{k_1+k_2},\quad&  g^{k_1}g^{-k_2}=g^{k_1-k_2}, \\
       g^{-k_1}g^{k_2}=g^{-k_1+k_2},\quad&  g^{-k_1}g^{-k_2}=g^{-k_1-k_2}.
\end{array}   
\end{align}
Two-point couplings of fields $\{\varphi_1,\varphi_2\}$ labeled by the conjugacy classes $\{g^{k_1},g^{k_2}\}$ in the 4D QFT obeys Eq. \eqref{eq:2_point}. 
After gauging the outer automorphism of $\mathbb{Z}_M$, we combine Eq.~(\ref{eq:ZM-rule}) and then we arrive at the following selection rules of the fields $\{\phi_1,\phi_2\}$ labeled by the class $\{[g^{k_1}],[g^{k_2}]\}$:
\begin{align}
\label{eq:2-p-1}
    [g^{k_1}][g^{k_2}] = [g^{k_1+k_2}] + [g^{M-k_1+k_2}].
\end{align}
It indicates that the two-point couplings of $\{\phi_1,\phi_2\}$ are allowed only when
\begin{align}
\label{eq:2-p-2}
    \pm k_1 \pm k_2=0 \quad (\mathrm{mod}\,M),
\end{align}
as calculated explicitly from Eq.~(\ref{eq:ZM-rule}), which is nothing but $[g^{k_1}]=[g^{k_2}]$. Hence, there are no mixing terms between fields labeled by different classes including their kinetic terms. 
However, in the case of three-point couplings including Yukawa couplings, one can realize the non-trivial structure due to the following selection rules for fields $\{\phi_1,\phi_2,\phi_3\}$ labeled by the class $\{[g^{k_1}],[g^{k_2}],[g^{k_3}]\}$:
\begin{align}
    [g^{k_1}][g^{k_2}][g^{k_3}] = [g^{k_1+k_2+k_3}] + [g^{M-k_1+k_2+k_3}]+ [g^{M+k_1-k_2+k_3}]+ [g^{M+k_1+k_2-k_3}].
\end{align}
Then, the three-point couplings are allowed only when
\begin{align}
    \pm k_1 \pm k_2 \pm k_3=0 \quad (\mathrm{mod}\,M).
    \label{eq:3-point-rule}
\end{align}
These are also derived explicitly by a calculation similar to Eqs.~(\ref{eq:ZM-rule}), (\ref{eq:2-p-1}), (\ref{eq:2-p-2}). 
One can extend this analysis for the selection rules of $n$ kinds of fields $\{\phi_1, ..., \phi_n\}$ labeled by the conjugacy classes $\{ [g^{k_1}],..., [g^{k_n}]\}$. 
From their selection rules:
\begin{align}
    [g^{k_1}]\cdots[g^{k_n}] = [g^{\sum_{i=1}^nk_i}] + [g^{M+\sum_j(\sum_{i=1,i\neq j}^n k_i-k_j)}],
\end{align}
it turns out that the $n$-point couplings are allowed only when
\begin{align}
    \sum_i \pm k_i =0\quad (\mathrm{mod}\,M).
\end{align}

%%%%%%%%%%%%%%%%%%%%%%%%%%%%%%%%%%%%%%%%%%%%%%%%%%%%%%%%%%%%%%%%%
\section{Yukawa textures for $M=3$}
\label{sec:M=3}
%%%%%%%%%%%%%%%%%%%%%%%%%%%%%%%%%%%%%%%%%%%%%%%%%%%%%%%%%%%%%%%%%

Here, we study Yukawa textures derived by our coupling selection rules for $M=3$. 
The fusion rules by $\mathbb{Z}_2$ gauging of $\mathbb{Z}_3$ are also known as the Fibonacci fusion rules.

\subsection{Textures by the Fibonacci fusion rules}
\label{sec:single}

When $M=3$, there are two classes $[g^0]$ and $[g^1]$.
These two classes are assigned to three generations of quarks and Higgs field(s).
If all three generations have the same class, 
all the entries of the $(3\times 3)$ Yukawa matrix are allowed, or the determinant of the Yukawa matrix is vanishing.
The latter case leads to at least one massless quark, which is not a realistic assignment.
The would-be realistic assignments are the following two patterns:
\begin{align}
\begin{array}{lll}
i&:&([g^0],[g^1],[g^1])\,, \notag \\ 
ii&:&([g^0],[g^0],[g^1])\,,
\end{array}
\end{align}
for three generations of quarks including their  permutations.
We combine these two for left-handed and right-handed quarks. 
In total, we have four possible combinations, which are shown in Table~\ref{tab:M=3}.
The second and third rows show assignments of classes for three generations of left- and right-handed quarks.
The last two rows show the Yukawa matrices when the Higgs field has the class $[g^0]$ and $[g^1]$, respectively.
Note that their permutations of rows and columns are possible by exchanging the ordering of $[g^0]$ and $[g^1]$ in three generations.
The Yukawa matrices in the case III and IV are transposed to each other.
The asterisks denote non-vanishing entries, which are allowed by our selection rule (\ref{eq:3-point-rule}).
The Yukawa matrices with vanishing determinants are denoted by $D=0$.
They lead to a massless quark.
Hereafter, we do not discuss such assignments.

When the Higgs field has the class $[g^0]$, 
the Yukawa textures for the flavor assignments I and II can be understood by 
the $\mathbb{Z}_2$ symmetry, where $[g^0]$ and $[g^1]$ correspond to 
$\mathbb{Z}_2$ even and odd, respectively. 
The three generations of both left-handed and right-handed quarks can correspond to a combination of one $\mathbb{Z}_2$ even mode and two odd modes.
The combination of two even modes and one odd mode leads to the same result.

When the Higgs field has the class $[g^1]$, 
the Yukawa textures with non-vanishing determinants in Table~\ref{tab:M=3} can not be derived by a group symmetry.

\begin{table}[h]
\begin{center}
\begin{tabular} {|c|c|c|c|c|} \hline 
Flavor & I &II& III & IV\\ \hline
left & $([g^0],[g^1],[g^1])$& $([g^0],[g^0],[g^1])$&$([g^0],[g^1],[g^1])$ & $([g^0],[g^0],[g^1])$  \\ \hline
right & $([g^0],[g^1],[g^1])$& $([g^0],[g^0],[g^1])$& $([g^0],[g^0],[g^1])$ &$([g^0],[g^1],[g^1])$ \\ \hline
Higgs $[g^0]$ & 
$
\begin{pmatrix}
* & 0 & 0 \\
0 & * & *  \\
0 & * & *
\end{pmatrix}
$
 &$
\begin{pmatrix}
* & * & 0 \\
* & * & 0  \\
0 & 0 & *
\end{pmatrix}
$&  $
\begin{pmatrix}
* & * & 0 \\
0 & 0 & *  \\
0 & 0 & *
\end{pmatrix}_{D=0}
$ & $
\begin{pmatrix}
* & 0 & 0 \\
* & 0 & 0  \\
0 & * & *
\end{pmatrix}_{D=0}
$
\\ \hline
Higgs $[g^1]$ &
$
\begin{pmatrix}
0 & * & * \\
* & * & *  \\
* & * & *
\end{pmatrix}
$
 & 
$
\begin{pmatrix}
0 & 0 & * \\
0 & 0 & *  \\
* & * & *
\end{pmatrix}_{D=0}
$
&  $
\begin{pmatrix}
0 & 0 & * \\
* & * & *  \\
* & * & *
\end{pmatrix}
$ 
&  $
\begin{pmatrix}
0 & * & * \\
0 & * & *  \\
* & * & *
\end{pmatrix}
$ 
\\ \hline
\end{tabular}
\caption{Yukawa matrices for $M=3$. The asterisks denote non-vanishing entries.}
\label{tab:M=3}
\end{center}
\end{table}

\subsection{Combinations of selection rules without group actions}
\label{sec:two_flavor}

We extend the previous analysis of the Fibonacci fusion rules to more general ones. For concreteness, suppose that there are two discrete symmetries, i.e., $\mathbb{Z}_2$ gauging of two different $\mathbb{Z}_M$ symmetries to control the Yukawa textures.
That is, the quarks as well as Higgs field(s) have their classes, i.e., $[g_1^k][g_2^m]$ ($k,m=0,1$).
These combinations of two discrete symmetries can lead to rich flavor structures. 
Indeed, the selection rule of the fields $\{\phi_1,\phi_2\}$ labeled by the class $\{[g^{k_1}][g^{m_1}],[g^{k_2}][g^{m_2}]\}$ is different from before:
\begin{align}
    [g^{k_1}][g^{m_1}]\cdot [g^{k_2}][g^{m_2}] &= [g^{k_1+k_2}][g^{m_1+m_2}] + [g^{M-k_1+k_2}][g^{m_1+m_2}]
    \nonumber\\
    &+[g^{k_1+k_2}][g^{M'-m_1+m_2}] + [g^{M-k_1+k_2}][g^{M'-m_1+m_2}] .
\end{align}
Hence, the two-point couplings of $\{\phi_1,\phi_2\}$ are allowed only when
\begin{align}
    \pm k_1 \pm k_2=0 \quad (\mathrm{mod}\,M),\quad 
    \mathrm{and}\quad
  \pm m_1 \pm m_2=0 \quad (\mathrm{mod}\,M'),
\end{align}
which is nothing but $[g^{k_1}]=[g^{k_2}]$ and $[g^{m_1}]=[g^{m_2}]$. Hence, there are no mixing terms between fields labeled by different classes including their kinetic terms.

For the selection rules of $n$ kinds of fields $\{\phi_1, ..., \phi_n\}$ labeled by the conjugacy classes $\{ [g^{k_1}][g^{m_1}],..., [g^{k_n}][g^{m_n}]\}$, their selection rule:
\begin{align}
    \left([g^{k_1}][g^{m_1}]\right)\cdots\left([g^{k_n}][g^{m_n}]\right) &= [g^{\sum_{i=1}^nk_i}][g^{\sum_{i=1}^nm_i}]+[g^{\sum_{i=1}^nk_i}][g^{M'+\sum_j(\sum_{i=1,i\neq j}^n m_i-m_j)}]
    \nonumber\\
    &+ [g^{M+\sum_j(\sum_{i=1,i\neq j}^n k_i-k_j)}][g^{\sum_{i=1}^nm_i}]
    \nonumber\\
    &+[g^{M+\sum_j(\sum_{i=1,i\neq j}^n k_i-k_j)}][g^{M'+\sum_j(\sum_{i=1,i\neq j}^n m_i-m_j)}],
\end{align}
lead to the $n$-point couplings are allowed only when
\begin{align}
    \sum_i \pm k_i =0\quad (\mathrm{mod}\,M),\quad 
    \mathrm{and}\quad
  \sum_i\pm m_i=0 \quad (\mathrm{mod}\,M').
\end{align}

Obviously, the flavor assignment with $k=m$ for all of the fields leads to the same results as the ones in Table~\ref{tab:M=3}.
Here, we do not study such cases. 
We study the case $M=M'=3$ in this section.

\subsubsection{Higgs sector with $[g_1^0][g_2^0]$}

First, we study the case that the Higgs field has the class $[g_1^0][g_2^0]$.
Since the flavor assignments III and IV are not 
realistic leading to a massless quark as shown in Table~\ref{tab:M=3}, we do not consider combinations including them.
On the other hand, the flavor assignments I and II lead to the same texture pattern up to permutations of matrix rows and columns.
As mentioned, the coupling selection rule by $[g^k]$ can be understood by the $\mathbb{Z}_2$ symmetry.
Thus, the selection rule by $[g_1^k][g_2^m]$ can correspond to the $\mathbb{Z}_2\times \mathbb{Z}'_2$ symmetry.
Suppose that the first $\mathbb{Z}_2$ corresponding to $[g^k_1]$ leads to the 
following Yukawa texture:
\begin{align}
    \begin{pmatrix}
* & 0 & 0 \\
0 & * & *  \\
0 & * & *
\end{pmatrix},
\label{eq:M3-1}
\end{align}
as shown in Table~\ref{tab:M=3}.
We may assign the $\ell$-th left-handed and $n$-th right-handed quarks to  the $\mathbb{Z}'_2$ even modes corresponding to $[g^m_2]$, and the others are $\mathbb{Z}'_2$ odd modes. 
Then, the $(\ell, n)$ entry in the above matrix is allowed, but 
the other entries in the $\ell$-th row and $n$-th column are ruled out.
For $(\ell,n)=(1,1)$, the same Yukawa texture (\ref{eq:M3-1}) remains to be allowed.
For example, when $(\ell,n)=(2,2)$, we can realize the following diagonal matrix:
\begin{align}
    \begin{pmatrix}
* & 0 & 0 \\
0 & * & 0  \\
0 & 0 & *
\end{pmatrix}.
\label{eq:M3-diagonal}
\end{align}
The other assignments of the class $[g_1^k][g_2^m]$ lead to 
its permutations of matrix rows and columns, or the Yukawa matrices with 
vanishing determinants.
Hence, the realistic texture is only the above diagonal matrix (\ref{eq:M3-diagonal}). 
We add these newly found textures in Table~\ref{tab:M=3-2}. Here, the underlines and wavy lines respectively represent a proper permutation leading to the textures which can not be produced in the previous analysis with the Fibonacci fusion rule. For instance, the texture \eqref{eq:M3-diagonal} can be realized by the following flavor assignments:
\begin{align}
\begin{array}{lll}
{\rm Left}&:&([g_1^0][g_2^1],~[g_1^1][g_2^0],~[g_1^1][g_2^1])\,, \notag \\
{\rm Right}&:&([g_1^0][g_2^1],~[g_1^1][g_2^0],~[g_1^1][g_2^1])\,.
\end{array}
\end{align}

\subsubsection{Higgs sector with $[g_1^0][g_2^1]$}

Next, we study the case that the Higgs field has the class $[g_1^0][g_2^1]$.
Similar to the previous case, the flavor assignment for $[g^k_1]$ must be 
I or II, and both lead to the same texture, i.e. eq.~(\ref{eq:M3-1}).

Let us study the flavor assignment I for $[g^m_2]$ including its possible permutations.
That can rule out one entry in the matrix (\ref{eq:M3-1}).
Then we can realize the following textures:
\begin{align}
\begin{pmatrix}
* & 0 & 0 \\
0 & 0 & *  \\
0 & * & *
\end{pmatrix}, 
\quad
\begin{pmatrix}
* & 0 & 0 \\
0 & * & 0  \\
0 & * & *
\end{pmatrix}, 
\quad
\begin{pmatrix}
* & 0 & 0 \\
0 & * & *  \\
0 & 0 & *
\end{pmatrix}, 
\quad
\begin{pmatrix}
* & 0 & 0 \\
0 & * & *  \\
0 & * & 0
\end{pmatrix}, 
\end{align}
as well as matrices with vanishing determinants.
Note that these four textures are 
related to each other by permutations of 
matrix rows and columns, when we adopt the flavor assignment III (IV) for $[g^m_2]$ including its possible permutations. 
That can rule out two entries in the same row (column) in the matrix so as to obtain the above textures as well as the matrices with vanishing determinants.
The above textures are only the possibilities for the Higgs sector with $[g_1^0][g_2^1]$. That is shown in Table~\ref{tab:M=3-2}.

\subsubsection{Higgs sector with $[g_1^1][g_2^1]$}

Here, we study the case that the Higgs field has the class $[g_1^1][g_2^1]$.
The flavor assignment II leads to a massless quark.
Thus, we study combinations of the flavor assignments I, III, IV for 
$[g^k_1][g^m_2]$.

When both the flavor assignments for $[g^k_1]$ and $[g^m_2]$ correspond to the assignment I, 
we can make the two-zero textures as 
\begin{align}
\begin{pmatrix}
0 & 0 & * \\
* & * & *  \\
* & * & *
\end{pmatrix},
\quad
\begin{pmatrix}
0 & * & * \\
* & 0 & *  \\
* & * & *
\end{pmatrix}, 
\quad
\begin{pmatrix}
0 & * & * \\
0 & * & *  \\
* & * & *
\end{pmatrix}, 
\end{align}
including the permutations of matrix rows and columns as well as the one-zero texture 
shown in Table~\ref{tab:M=3} including 
the permutations of rows and columns.

When combinations of I and III are used for $[g^k_1]$ and $[g^m_2]$, we obtain the following textures:
\begin{align}
\begin{pmatrix}
0 & 0 & * \\
0 & * & *  \\
* & * & *
\end{pmatrix}, 
\quad
\begin{pmatrix}
0 & 0 & * \\
* & * & 0  \\
* & * & *
\end{pmatrix}, 
\end{align}
including permutations of matrix rows and columns 
as well as the matrices with vanishing determinants.
Similarly, combinations of I and IV are used for $[g^k_1]$ and $[g^m_2]$, we obtain the following textures:
\begin{align}
\begin{pmatrix}
0 & 0 & * \\
0 & * & *  \\
* & * & *
\end{pmatrix}, 
\quad
\begin{pmatrix}
0 & * & * \\
0 & * & * \\
* & 0 & *
\end{pmatrix}, 
\end{align}
including permutations of matrix rows and columns 
as well as the matrices with vanishing determinants.

When both the flavor assignments for $[g^k_1]$ and $[g^m_2]$ correspond to the assignment III, we can construct the textures, whose two rows have two zeros, i.e.,
\begin{align}
\begin{pmatrix}
0 & 0 & * \\
* & * & *  \\
* & 0 & 0
\end{pmatrix},
\end{align}
including permutations of rows and columns as well as matrices with vanishing determinants.
Similarly, when both the flavor assignments for $[g^k_1]$ and $[g^m_2]$ correspond to the assignment IV, we can construct the textures, whose two columns have two zeros, i.e.,
\begin{align}
\begin{pmatrix}
0 & * & * \\
0 & * & 0  \\
* & * & 0
\end{pmatrix},
\end{align}
including permutations of rows and columns as well as matrices with vanishing determinants.

When combinations of III and IV are used for $[g^k_1]$ and $[g^m_2]$, we obtain the textures, where one row and one column have two zeros, 
i.e.,
\begin{align}
\begin{pmatrix}
0 & 0 & * \\
* & * & 0  \\
* & * & 0
\end{pmatrix},
\end{align}
including permutations of rows and columns.
Also combinations of III and IV can lead to the following 
texture:
\begin{align}
\begin{pmatrix}
0 & 0 & * \\
0 & * & *  \\
* & * & *
\end{pmatrix}
.
\end{align}
Other combinations lead to matrices with vanishing determinants.

The above results are summarized in Table~\ref{tab:M=3-2}.

{\footnotesize
\begin{longtable}{|c|c|c|c|}
\caption{Yukawa texture for $M=3$, where the underlines and wavy lines respectively represent a proper permutation leading to the textures which can not be produced in the previous analysis with the Fibonacci fusion rule.}
\label{tab:M=3-2}\\ \hline
Flavor ($[g^k_1],[g^m_2]$) & Higgs $[g_1^0][g_2^0]$ & Higgs $[g_1^0][g_2^1]$ &     Higgs $[g_1^1][g_2^1]$  \\ \hline
\begin{tabular}{l}
(I, I) : \\
Left\,\,\,\,\,:  $(\underline{[g_1^0]}\uwave{[g_2^0]},\underline{[g_1^1]}\uwave{[g_2^1]},\underline{[g_1^1]}\uwave{[g_2^1]})$\\ 
Right\,: 
$(\underline{[g_1^0]}\uwave{[g_2^0]},\underline{[g_1^1]}\uwave{[g_2^1]},\underline{[g_1^1]}\uwave{[g_2^1]})$
\end{tabular}
& $
    \begin{pmatrix}
* & 0 & 0 \\
0 & * & 0  \\
0 & 0 & *
\end{pmatrix} $ & $
\begin{pmatrix}
* & 0 & 0 \\
0 & * & 0  \\
0 & * & *
\end{pmatrix}
$ & $
\begin{pmatrix}
0 & 0 & * \\
* & * & *  \\
* & * & *
\end{pmatrix}, 
\begin{pmatrix}
0 & * & * \\
* & 0 & *  \\
* & * & *
\end{pmatrix}, 
\begin{pmatrix}
0 & * & * \\
0 & * & *  \\
* & * & *
\end{pmatrix}
$ \\ \hline
\begin{tabular}{l}
(I, II) : \\
Left\,\,\,\,\,: $(\underline{[g_1^0]}\uwave{[g_2^0]},\underline{[g_1^1]}\uwave{[g_2^0]},\underline{[g_1^1]}\uwave{[g_2^1]})$\\ 
Right\,: 
$(\underline{[g_1^0]}\uwave{[g_2^0]},\underline{[g_1^1]}\uwave{[g_2^0]},\underline{[g_1^1]}\uwave{[g_2^1]})$\\
and (II, II) :\\
Left\,\,\,\,\,: $(\underline{[g_1^0]}\uwave{[g_2^0]},\underline{[g_1^0]}\uwave{[g_2^0]},\underline{[g_1^1]}\uwave{[g_2^1]})$\\ 
Right\,: 
$(\underline{[g_1^0]}\uwave{[g_2^0]},\underline{[g_1^0]}\uwave{[g_2^0]},\underline{[g_1^1]}\uwave{[g_2^1]})$
\end{tabular}
& $
    \begin{pmatrix}
* & 0 & 0 \\
0 & * & 0  \\
0 & 0 & *
\end{pmatrix} 
$ & $D=0$ & $D=0$ \\ \hline
\begin{tabular}{l}
(I, III) : \\
Left\,\,\,\,\,:  $(\underline{[g_1^0]}\uwave{[g_2^0]},\underline{[g_1^1]}\uwave{[g_2^1]},\underline{[g_1^1]}\uwave{[g_2^1]})$\\ 
Right\,: 
$(\underline{[g_1^0]}\uwave{[g_2^0]},\underline{[g_1^1]}\uwave{[g_2^0]},\underline{[g_1^1]}\uwave{[g_2^1]})$
\end{tabular}
& $D=0$ & $
\begin{pmatrix}
* & 0 & 0 \\
0 & * & 0  \\
0 & * & *
\end{pmatrix}
$ & $
\begin{pmatrix}
0 & 0 & * \\
0 & * & *  \\
* & * & *
\end{pmatrix}, 
\begin{pmatrix}
0 & 0 & * \\
* & * & 0  \\
* & * & *
\end{pmatrix}
$ \\ \hline
\begin{tabular}{l}
(I, IV) : \\
Left\,\,\,\,\,:  $(\underline{[g_1^0]}\uwave{[g_2^0]},\underline{[g_1^1]}\uwave{[g_2^0]},\underline{[g_1^1]}\uwave{[g_2^1]})$\\ 
Right\,: 
$(\underline{[g_1^0]}\uwave{[g_2^0]},\underline{[g_1^1]}\uwave{[g_2^1]},\underline{[g_1^1]}\uwave{[g_2^1]})$
\end{tabular} & $D=0$ & $
\begin{pmatrix}
* & 0 & 0 \\
0 & * & 0  \\
0 & * & *
\end{pmatrix}
$ & $
\begin{pmatrix}
0 & 0 & * \\
0 & * & *  \\
* & * & *
\end{pmatrix}, 
\begin{pmatrix}
0 & * & * \\
0 & * & *  \\
* & 0 & *
\end{pmatrix}
$ \\ \hline
\begin{tabular}{l}
(II, I) : \\
Left\,\,\,\,\,:  $(\underline{[g_1^0]}\uwave{[g_2^0]},\underline{[g_1^0]}\uwave{[g_2^1]},\underline{[g_1^1]}\uwave{[g_2^1]})$\\ 
Right\,: 
$(\underline{[g_1^0]}\uwave{[g_2^0]},\underline{[g_1^0]}\uwave{[g_2^1]},\underline{[g_1^1]}\uwave{[g_2^1]})$
\end{tabular} & $
    \begin{pmatrix}
* & 0 & 0 \\
0 & * & 0  \\
0 & 0 & *
\end{pmatrix} $ & $
\begin{pmatrix}
* & 0 & 0 \\
0 & * & 0  \\
0 & * & *
\end{pmatrix}
$ & $D=0$ \\ \hline
\begin{tabular}{l}
(II, III) : \\
Left\,\,\,\,\,: $(\underline{[g_1^0]}\uwave{[g_2^0]},\underline{[g_1^0]}\uwave{[g_2^1]},\underline{[g_1^1]}\uwave{[g_2^1]})$\\ 
Right\,: 
$(\underline{[g_1^0]}\uwave{[g_2^0]},\underline{[g_1^0]}\uwave{[g_2^0]},\underline{[g_1^1]}\uwave{[g_2^1]})$\\
and (II, IV) :\\
Left\,\,\,\,\,: $(\underline{[g_1^0]}\uwave{[g_2^0]},\underline{[g_1^0]}\uwave{[g_2^0]},\underline{[g_1^1]}\uwave{[g_2^1]})$\\ 
Right\,: 
$(\underline{[g_1^0]}\uwave{[g_2^0]},\underline{[g_1^0]}\uwave{[g_2^1]},\underline{[g_1^1]}\uwave{[g_2^1]})$
\end{tabular} & $D=0$ & $
\begin{pmatrix}
* & 0 & 0 \\
0 & * & 0  \\
0 & * & *
\end{pmatrix}
$ & $D=0$ \\ \hline 
\begin{tabular}{l}
(III, I) : \\
Left\,\,\,\,\,:  $(\underline{[g_1^0]}\uwave{[g_2^0]},\underline{[g_1^1]}\uwave{[g_2^1]},\underline{[g_1^1]}\uwave{[g_2^1]})$\\ 
Right\,: 
$(\underline{[g_1^0]}\uwave{[g_2^0]},\underline{[g_1^0]}\uwave{[g_2^1]},\underline{[g_1^1]}\uwave{[g_2^1]})$
\end{tabular} & $D=0$ & $D=0$ &  $
\begin{pmatrix}
0 & 0 & * \\
0 & * & *  \\
* & * & *
\end{pmatrix}, 
\begin{pmatrix}
0 & 0 & * \\
* & * & 0  \\
* & * & *
\end{pmatrix}
$

\\ \hline
\begin{tabular}{l}
(III, II) : \\
Left\,\,\,\,\,:  $(\underline{[g_1^0]}\uwave{[g_2^0]},\underline{[g_1^1]}\uwave{[g_2^0]},\underline{[g_1^1]}\uwave{[g_2^1]})$\\ 
Right\,: 
$(\underline{[g_1^0]}\uwave{[g_2^0]},\underline{[g_1^0]}\uwave{[g_2^0]},\underline{[g_1^1]}\uwave{[g_2^1]})$
\end{tabular} & $D=0$ & $D=0$ &  $D=0$  \\ \hline
 \begin{tabular}{l}
(III, III) : \\
Left\,\,\,\,\,:  $(\underline{[g_1^0]}\uwave{[g_2^0]},\underline{[g_1^1]}\uwave{[g_2^1]},\underline{[g_1^1]}\uwave{[g_2^1]})$\\ 
Right\,: 
$(\underline{[g_1^0]}\uwave{[g_2^0]},\underline{[g_1^0]}\uwave{[g_2^0]},\underline{[g_1^1]}\uwave{[g_2^1]})$
\end{tabular} &$D=0$ & $D=0$ & $
 \begin{pmatrix}
0 & 0 & * \\
* & * & *  \\
* & 0 & 0
\end{pmatrix}
 $\\ \hline
 \begin{tabular}{l}
(III, IV) : \\
Left\,\,\,\,\,:  $(\underline{[g_1^0]}\uwave{[g_2^0]},\underline{[g_1^1]}\uwave{[g_2^0]},\underline{[g_1^1]}\uwave{[g_2^1]})$\\ 
Right\,: 
$(\underline{[g_1^0]}\uwave{[g_2^0]},\underline{[g_1^0]}\uwave{[g_2^1]},\underline{[g_1^1]}\uwave{[g_2^1]})$
\end{tabular} &$D=0$ & $D=0$ &  $
 \begin{pmatrix}
0 & 0 & * \\
* & * & 0  \\
* & * & 0
\end{pmatrix},
\begin{pmatrix}
0 & 0 & * \\
0 & * & *  \\
* & * & *
\end{pmatrix}
$  \\ \hline
 \begin{tabular}{l}
(IV, I) : \\
Left\,\,\,\,\,:  $(\underline{[g_1^0]}\uwave{[g_2^0]},\underline{[g_1^0]}\uwave{[g_2^1]},\underline{[g_1^1]}\uwave{[g_2^1]})$\\ 
Right\,: 
$(\underline{[g_1^0]}\uwave{[g_2^0]},\underline{[g_1^1]}\uwave{[g_2^1]},\underline{[g_1^1]}\uwave{[g_2^1]})$
\end{tabular} & $D=0$ & $D=0$ &   $
\begin{pmatrix}
0 & 0 & * \\
0 & * & *  \\
* & * & *
\end{pmatrix}, 
\begin{pmatrix}
0 & * & * \\
0 & * & *  \\
* & 0 & *
\end{pmatrix}
$   \\ \hline
 \begin{tabular}{l}
(IV, II) : \\
Left\,\,\,\,\,:  $(\underline{[g_1^0]}\uwave{[g_2^0]},\underline{[g_1^0]}\uwave{[g_2^0]},\underline{[g_1^1]}\uwave{[g_2^1]})$\\ 
Right\,: 
$(\underline{[g_1^0]}\uwave{[g_2^0]},\underline{[g_1^1]}\uwave{[g_2^0]},\underline{[g_1^1]}\uwave{[g_2^1]})$
\end{tabular} & $D=0$ & $D=0$ &  $D=0$  \\ \hline
 \begin{tabular}{l}
(IV, III) : \\
Left\,\,\,\,\,:  $(\underline{[g_1^0]}\uwave{[g_2^0]},\underline{[g_1^0]}\uwave{[g_2^1]},\underline{[g_1^1]}\uwave{[g_2^1]})$\\ 
Right\,: 
$(\underline{[g_1^0]}\uwave{[g_2^0]},\underline{[g_1^1]}\uwave{[g_2^0]},\underline{[g_1^1]}\uwave{[g_2^1]})$
\end{tabular} &$D=0$ & $D=0$ & $
 \begin{pmatrix}
0 & * & * \\
0 & * & *  \\
* & 0 & 0
\end{pmatrix}, 
\begin{pmatrix}
0 & 0 & * \\
0 & * & *  \\
* & * & *
\end{pmatrix}
$ \\ \hline
 \begin{tabular}{l}
(IV, IV) : \\
Left\,\,\,\,\,:  $(\underline{[g_1^0]}\uwave{[g_2^0]},\underline{[g_1^0]}\uwave{[g_2^0]},\underline{[g_1^1]}\uwave{[g_2^1]})$\\ 
Right\,: 
$(\underline{[g_1^0]}\uwave{[g_2^0]},\underline{[g_1^1]}\uwave{[g_2^1]},\underline{[g_1^1]}\uwave{[g_2^1]})$
\end{tabular} &$D=0$ & $D=0$ & $
 \begin{pmatrix}
0 & * & * \\
0 & * & 0  \\
* & * & 0
\end{pmatrix}$
\\ \hline
\end{longtable}
}

\subsection{Up and down sectors}

Here, we study combinations of  Yukawa matrices in the up and down sectors by use of the results in the previous section.
Each matrix corresponds to one in Table~\ref{tab:M=3-2} including their permutations.
There are nine real observables including six quark masses and three mixing angles as well as one phase.
Totally, we need at least nine non-vanishing entries in the up and down Yukawa matrices.
In addition, each of the up and down Yukawa matrices must have at least three non-vanishing entries to realize all of the three quark masses.
Under these conditions, we study the up and down textures, which include more zeros.

\subsubsection{Same texture for up and down sectors}

First, we study combinations of the up and down Yukawa matrices, where their textures are the same for the up and down sectors.
As mentioned above, we need at least nine non-vanishing entries totally in Yukawa matrices of the up and down sectors.
Four-zero textures for both the up and down sectors include minimal parameters.
Three-zero textures are next-to-minimal ones.

One of the four-zero textures is 
\begin{align}
(a)\qquad Y_{u}, \, Y_{d}=
\begin{pmatrix}
* & * & 0 \\
0 & * & 0  \\
0 & * & *
\end{pmatrix}.
\label{(a)}
\end{align}
For example, this texture is realized by the following flavor assignments:
\begin{align}
\begin{array}{lll}
{\rm Left}&:&([g^1_1][g^0_2],~[g^0_1][g^0_2],~[g^0_1][g^1_2])\,, \notag \\
{\rm Right}&:&([g^0_1][g^1_2],~[g^1_1][g^1_2],~[g^1_1][g^0_2])\,,
\end{array}
\end{align}
for both the up and down sectors when 
the Higgs field has $[g^1_1][g^1_2]$. 
They are a permutation of the flavor assignment (IV,IV) in Table \ref{tab:M=3-2} and can lead to all of the non-vanishing masses and mixing angles.
When we exchange the assignments of left- and right-handed quarks, we obtain its transposed matrix,
\begin{align}
(b)\qquad Y_{u}, \, Y_{d}=
\begin{pmatrix}
* & 0 & 0 \\
* & * & *  \\
0 & 0 & *
\end{pmatrix},
\label{(b)}
\end{align}
corresponding to a permutation of the flavor assignment (III,III) in Table~\ref{tab:M=3-2}.
They also lead to all of the non-vanishing masses and mixing angles.

The other four-zero texture is 
\begin{align}
\begin{pmatrix}
* & 0 & 0 \\
0 & * & *  \\
0 & * & *
\end{pmatrix},
\end{align}
corresponding to a permutation of the flavor assignment (III,IV) when the Higgs field has $[g^1_1][g^1_2]$.
However, if both the up and down sectors correspond to the same texture as the above, one can not realize all of three mixing angles.
Hence, this possibility is ruled out.

One of three-zero textures is 
\begin{align}
(c) \qquad Y_{u},\, Y_{d}=
\begin{pmatrix}
0 & 0 & * \\
0 & * & *  \\
* & * & *
\end{pmatrix}.
\label{(c)}
\end{align}
For example, this texture is realized by the following flavor assignments:
\begin{align}
\begin{array}{lll}
{\rm Left}&:&([g^0_1][g^0_2],~[g^1_1][g^0_2],~[g^1_1][g^1_2])\,, \notag \\
{\rm Right}&:&([g^0_1][g^0_2],~[g^0_1][g^1_2],~[g^1_1][g^1_2])\,,
\end{array}
\end{align}
for both the up and down sectors corresponding to the flavor assignment (III,IV) when the Higgs field has $[g^1_1][g^1_2]$. 
Other assignments can also lead to this texture as shown in Table~\ref{tab:M=3-2}.

Another three-zero texture is 
\begin{align}
\begin{pmatrix}
0 & 0 & * \\
* & * & 0  \\
* & * & *
\end{pmatrix}.
\end{align}
For example, this texture is realized by the following flavor assignments:
\begin{align}
\begin{array}{lll}
{\rm Left}&:&([g^0_1][g^1_2],~[g^1_1][g^0_2],~[g^1_1][g^1_2])\,, \notag \\
{\rm Right}&:&([g^0_1][g^1_2],~[g^0_1][g^1_2],~[g^1_1][g^0_2])\,,
\end{array}
\end{align}
corresponding to the flavor assignment (I,III) when the Higgs field has $[g^1_1][g^1_2]$. Its transposed matrix,
\begin{align}
\begin{pmatrix}
0 & * & * \\
0 & * & *  \\
* & 0 & *
\end{pmatrix}.
\end{align}
can be obtained when we replace III with IV. 
This texture was shown in Ref.~\cite{Tanimoto:2016rqy} as the quark mass matrix as the down sector.

These three-zero textures for both the up and down sectors include 
more free parameters.
In this sense, three-zero textures for both the up and down sectors are not phenomenologically 
interesting.
However, the texture (\ref{(c)}) has the specific property that it has a non-trivial CP phase leading to the weak CP, but such CP phase does not appear in its determinant.

\subsubsection{Up and down different textures}

Here, we study the Yukawa textures, whose structures are different between the up and down sectors.
As mentioned before, the minimal textures correspond to 
the up and down Yukawa matrices, which have a totally nine non-vanishing entries.
For either up or down sector, the minimal one is the diagonal matrix.
The other must have at least six non-vanishing entries.
Thus, one of the minimal textures is 
\begin{align}
(d) \qquad Y_{u}=
\begin{pmatrix}
* & 0 & 0 \\
0 & * & 0  \\
0 & 0 & *
\end{pmatrix}
,\qquad Y_{d}=
\begin{pmatrix}
0 & 0 & * \\
0 & * & *  \\
* & * & *
\end{pmatrix}.
\label{(d)}
\end{align}
For example, the texture (\ref{(d)}) can be realized by the following flavor assignment:
\begin{align}
\begin{array}{lll}
{\rm Left}&:&([g^0_1][g^0_2],~[g^1_1][g^0_2],~[g^1_1][g^1_2])\,, \notag \\
{\rm Right\,(up-sector)}&:&([g^0_1][g^0_2],~[g^1_1][g^0_2],~[g^1_1][g^1_2])\,, \\
{\rm Right\,(down-sector)}&:&([g^0_1][g^0_2],~[g^0_1][g^1_2],~[g^1_1][g^1_2])\,,
\notag
\end{array}
\end{align}
when the up-sector Higgs has $[g^0_1][g^0_2]$ and the down-sector Higgs has 
$[g^1_1][g^1_2]$.

Here, we give a comment on the Higgs sector.
If the textures can be realized by the Higgs fields with the same class $[g^k_1][g^m_2]$ 
for both the up and down sectors, the Higgs sector can correspond to a single field like the Standard Model.
If they have different classes, such models can be realized, e.g., as (non-supersymmetric) type II two-Higgs doublet model and supersymmetric extension to the Standard Model.
If their classes are different in the supersymmetric model, the $\mu$-term is not allowed.
We have to introduce a singlet field $S$ with a proper class like the next-to-minimal supersymmetric standard model.

\section{Textures for $M=4$ and $5$}
\label{sec:M=4-5}

Similarly, we can study the possible textures for $M=4$ and 5.

\subsection{$M=4$}
\label{sec:M=4}

For $M=4$, there are three classes, $[g^0],[g^1],[g^2]$.
Combinations of allowed Yukawa couplings are as follows:
\begin{align}
[g^0][g^0][g^0], \qquad [g^0][g^1][g^1], \qquad [g^0][g^2][g^2],  \qquad [g^1][g^1][g^2].
\end{align}

If all three generations of the left-handed or right-handed quarks are assigned to the same class, all the entries in the Yukawa matrix are allowed, or the determinant of the Yukawa matrix becomes vanishing.
The other possible assignments including their permutations are shown in Table~\ref{tab:M=4-1}.

\begin{table}[h]
\centering
\scalebox{0.89}{
\begin{tabular}{|c|c|c|c|c|c|c|c|}\hline
Flavor & i & ii & iii & iv & v&  vi & vii \\ \hline
& $[g^0][g^1][g^2]$ & $[g^0][g^0][g^1]$ & $[g^0][g^0][g^2]$ & $[g^0][g^1][g^1]$ & $[g^0][g^2][g^2]$ &$[g^1][g^1][g^2]$ &$[g^1][g^2][g^2]$ \\ \hline
\end{tabular}
}
\caption{Possible assignments for three generations when $M=4$.}
\label{tab:M=4-1}
\end{table}

We combine these possible assignments for the left- and right-handed quarks.
Results are shown in Appendix \ref{app:M=4}.
Many of them lead to the Yukawa matrices with vanishing determinants and a massless quark.
Matrices with non-vanishing determinants are as follows:
\begin{align}
\begin{pmatrix}
*&0&0\\
0&*&0\\
0&0&*
\end{pmatrix}, \quad
\begin{pmatrix}
*&*&0\\
*&*&0\\
0&0&*
\end{pmatrix},
\end{align}
including permutations of rows and columns.
These have been already obtained in Table~\ref{tab:M=3} in section \ref{sec:M=3}.
If the up or down sector corresponds to the diagonal matrix, we have no sufficient number of parameters.
Thus, the following combination from tables in Appendix \ref{app:M=4}:
\begin{align}
    Y_u=
\begin{pmatrix}
*&*&0\\
*&*&0\\
0&0&*
\end{pmatrix}, \qquad 
    Y_d=
\begin{pmatrix}
0&0&*\\
*&*&0\\
*&*&0
\end{pmatrix},
\end{align}
including permutations has more free parameters, but this leads to not being realistic among mixing angles.

\subsection{$M=5$}
\label{sec:M=5}

For $M=5$, there are three classes, $[g^0],[g^1],[g^2]$.
Combinations of allowed Yukawa couplings are as follows:
\begin{align}
[g^0][g^0][g^0], \qquad [g^0][g^1][g^1], \qquad [g^0][g^2][g^2],  \qquad [g^1][g^1][g^2], \qquad [g^1][g^2][g^2].
\end{align}

By the same reason as $M=3$ and 4, we do not consider the flavor assignments, where all of the three generations correspond to the same class.
All the possible assignments including their permutations are shown in Table~\ref{tab:M=5-1}.

\begin{table}[h]
\centering
\scalebox{0.9}{
\begin{tabular}{|c|c|c|c|c|c|c|c|}\hline
Flavor & i & ii & iii & iv & v&  vi & vii    \\ \hline
& $[g^0][g^1][g^2]$ & $[g^0][g^0][g^1]$ & $[g^0][g^0][g^2]$ & $[g^0][g^1][g^1]$ & $[g^0][g^2][g^2]$ &$[g^1][g^1][g^2]$ &$[g^1][g^2][g^2]$  \\ \hline
\end{tabular}
}
\caption{Possible assignments for three generations when $M=5$.}
\label{tab:M=5-1}
\end{table}

We combine these possible assignments for the left- and right-handed quarks.
Results are shown in Appendix \ref{app:M=5}.
Many of them lead to the Yukawa matrices with vanishing determinants and a massless quark.
Matrices with non-vanishing determinants are as follows:
\begin{align}
\begin{pmatrix}
0&*&0\\
*&0&*\\
0&*&*
\end{pmatrix}, \quad
\begin{pmatrix}
0&0&*\\
0&*&*\\
*&*&0
\end{pmatrix}, 
\label{eq:M=5-1}
\end{align}
\begin{align}
\begin{pmatrix}
*&0&0\\
0&*&0\\
0&0&*
\end{pmatrix}, \quad
\begin{pmatrix}
0&0&*\\
*&*&*\\
*&*&0
\end{pmatrix}, \quad 
\begin{pmatrix}
*&*&0\\
*&*&0\\
0&0&*
\end{pmatrix}, \quad
\begin{pmatrix}
*&*&0\\
*&*&0\\
*&*&*
\end{pmatrix}, \quad
\begin{pmatrix}
0&*&*\\
*&*&*\\
*&*&*
\end{pmatrix},
\label{eq:M=5-2}
\end{align}
including permutations of rows and columns.
The textures in Eq.~(\ref{eq:M=5-2}) are already obtained for $M=3$.
The textures in Eq.~(\ref{eq:M=5-1}) are new ones.

In particular, the first one in Eq.~\eqref{eq:M=5-1} is 
so-called {\it the nearest neighbor interaction} (NNI) form,
which is considered as a 
“general” form of both up- and down-types quark mass matrices because this form is achieved
 by the transformation that leaves the left- handed gauge interaction invariant \cite{Branco:1988iq}.
Thus, one can realize the realistic quark masses and mixing angles as well as the CP phase by 
\begin{align}
    (e) \qquad Y_u, Y_d=
    \begin{pmatrix}
0&*&0\\
*&0&*\\
0&*&*
\end{pmatrix}.
\label{(f)}
\end{align}
For example, this texture can be realized by assigning $([g^0],[g^1],[g^2])$ to three generations of quarks in both the up and down sectors when the Higgs field have $[g^1]$ as shown in Ref.~\cite{Kobayashi:2024cvp}. 
This texture includes non-vanishing entries more than the minimal one.
We can combine this with the selection rule for $M=3$ so as to forbid one of entries.
For example, the following texture:
\begin{align}
    (f) \qquad Y_u= 
 \begin{pmatrix}
0&*&0\\
*&0&*\\
0&0&*
\end{pmatrix},
\quad
    Y_d=
    \begin{pmatrix}
0&*&0\\
*&0&*\\
0&*&*
\end{pmatrix}
\label{(g)}
\end{align}
would be interesting.
For example, for $M=3$, we use the following assignment:
\begin{align}
\begin{array}{lll}
{\rm Left}&: &([g^1_2],~[g^1_2],~[g^0_2])\,, \notag \\
{\rm Right\,(up-sector)}&: &([g^1_2],~[g^0_2],~[g^1_2])\,,
\\
{\rm Right\,(down-sector)}&:& ([g^0_2],~[g^1_2],~[g^1_2])\,,
\notag
\end{array}
\end{align}
in addition to the flavor assignment leading to Eq.~(\ref{(f)}) 
when the Higgs field has $[g^1_2]$ for $M=3$.

When the second one in Eq.~\eqref{eq:M=5-1} is included in the either up sector or down sector, one can not derive realistic results.

%%%%%%%%%%%%%%%%%%%%%%%%%%%%%%%%%%%%%%%%%%%%%%%%%%%%%%%%%%%%%%%%%
%%%%%%%%%%%%%%%%%%%%%%%%%%%%%%%%%%%%%%%%%%%%%%%%%%%%%%%%%%%%%%%%%
\section{Phenomenology of Models}
\label{sec:phenomenology}
In this section, we discuss the phenomenological implication
of interesting textures presented in  sections \ref{sec:M=3} and 
 \ref{sec:M=4-5}.
 
%%%%%%%%%%%%%%%%%%%%%%%%%%%%%%%
\subsection{Phase structure of texture { $(e)$}}
\label{phase-structure}

In order to understand the phase structure of texture zeros in general, we begin with discussing the NNI type  matrix in Eq.~\eqref{(f)},
which is consistent with experimental data.
The Yukawa matrices of  quarks are: 
\begin{align}
&  Y_q =
\begin{pmatrix}
0  &  A_q &0  \\
A'_q & 0 & B_Q \\
0& B'_q& C_q
\end{pmatrix}_{LR}\,,
\label{NNI-quark}
\end{align}
where  the  coefficient of each element is complex in general
\footnote{Among ten phases, eight phases  are removed  
	by the redefinition of the quark fields. }. 

We write down the NNI Yukawa matrix of quarks
including phase factors explicitly as:
\begin{align}
&  Y_q =
\begin{pmatrix}
0  &  a_q\, e^{i\phi_{a_q}} &0  \\
a'_q \, e^{i\phi_{a'_q}} & 0 & b_q \, e^{i\phi_{b_q}} \\
0& b'_q  \, e^{i\phi_{b'_q}}& c_q  \, e^{i\phi_{c_q}}
\end{pmatrix}_{LR}\,,
\label{NNI-phase1}
\end{align}
where coefficients $a_q$, $a'_q$, $b_q$, $b'_q$ and $c_q$ are real.

All phase factors are removed
in the Yukawa matrices by multiplying the phase matrices
 as follows:
 \begin{align}
 &  Y_q = P_{Lq}
 \begin{pmatrix}
 0  &  a_q &0  \\
 a'_q  & 0 & b_q \\
 0& b'_q & c_q 
 \end{pmatrix}\, P_{Rq}^*\,,
 \label{NNI-phase2}
 \end{align}
 where
 \begin{align}
 P_{Lq}=
 \begin{pmatrix}
e^{i(\phi_{a_q}-\phi_{b'_q})}  & 0&0  \\
0 & e^{i(\phi_{b_q}-\phi_{c_q})} &0 \\
 0& 0 & 1
 \end{pmatrix}\,,\qquad 
  P_{Rq}^*=
 \begin{pmatrix}
 e^{i(\phi_{a'_q}-\phi_{b_q}+\phi_{c_q})}  & 0&0  \\
 0 & e^{i\phi_{b'_q}} &0 \\
 0& 0 & e^{i\phi_{c_q}}
 \end{pmatrix}\,.
 \label{phase-NNI}
 \end{align}
 The CP phase of the CKM matrix comes from the left-handed
 phase matrix $P_{Lq}$.
In the case of   the NNI down- and up-type quark mass matrices,  the observed CP violation is given by
two parameters $\sigma$ and $\tau$ of  the phase matrix 
 \begin{align}
P_{L}=
\begin{pmatrix}
e^{i\sigma} & 0&0  \\
0 & e^{i\tau} &0 \\
0& 0 & 1
\end{pmatrix}\,,
\label{CKM-phase}
\end{align}
which is the combination of both down- and up-type phase matrices.

%%%%%%%%%%%%%%%%%%%%%%%%%%%%%%
\subsection{Model $(f)$}
We consider the non-trivial Yukawa matrices, which are derived in sections 3 and 4, and 
examine whether they work well or not.
At first, we discuss the texture in Eq.~\eqref{(g)}, which  is explicitly written as:
 \begin{align}
 &  Y_q =
 \begin{pmatrix}
 0  &  a_q\, e^{i\phi_{a_q}} &0  \\
 a'_q \, e^{i\phi_{a'_q}} & 0 & b_q \, e^{i\phi_{b_q}} \\
 0& 0& c_Q  \, e^{i\phi_{c_q}}
 \end{pmatrix}_{LR}\,,
 \label{NNI-like-phase1}
 \end{align}
 where the (3,2) entry vanishes.
All phase factors are also removed
in the Yukawa matrices by multiplying the phase matrices
as follows:
\begin{align}
&  Y_q = P_{Lq}
\begin{pmatrix}
0  &  a_q &0  \\
a'_q  & 0 & b_q \\
0& 0 & c_q 
\end{pmatrix}\, P_{Rq}^*
\,,
\label{NNI-like-phase2}
\end{align}
with
\begin{align}
P_{Lq}=
\begin{pmatrix}
1  & 0&0  \\
0 & e^{i(\phi_{b_q}-\phi_{c_q})} &0 \\
0& 0 & 1
\end{pmatrix}\,,\qquad 
P_{Rq}^*=
\begin{pmatrix}
e^{i(\phi_{a'_q}-\phi_{b_q}+\phi_{c_q})}  & 0&0  \\
0 & e^{i\phi_{a_q}} &0 \\
0& 0 & e^{i\phi_{c_q}}
\end{pmatrix}\,.
\label{phase-NNI-like2}
\end{align}

 %%%%%%%%%%%%%%%%%%%%%%%%%%%

Let us discuss a model of the quark mass matrix,
where the down-type Yukawa matrix is the NNI type, on the other hand, the up-type one is the NNI-like ($b'_u=0$) such as 
\begin{align}
&  Y_d =
\begin{pmatrix}
0  &  A_d &0  \\
A'_d & 0 & B_d \\
0& B'_d& C_d
\end{pmatrix}_{LR}\,,
\qquad
Y_u =
\begin{pmatrix}
0  &  A_u &0  \\
A'_u & 0 & B_u \\
0& 0 & C_u
\end{pmatrix}_{LR}\,,
\label{model-1}
\end{align}
where each coefficient of the matrices is complex.
The model is given by ten parameters,
which are adjusted by ten observables.
The ten parameters are given as follows.
The down-type Yukawa matrix is given by the two phase matrices and
real Yukawa matrix:
\begin{align}
&  Y_d = P_{Ld}
\hat Y_d\, P_{Rd}^*\,,\qquad 
  \hat Y_d = 
\begin{pmatrix}
0  &  a_d &0  \\
a'_d  & 0 & b_d \\
0& b'_d & c_d 
\end{pmatrix}\,,
\label{down-matrix}
\end{align}
\begin{align}
P_{Ld}=
\begin{pmatrix}
e^{i(\phi_{a_d}-\phi_{b'_d})}  & 0&0  \\
0 & e^{i(\phi_{b_d}-\phi_{c_d})} &0 \\
0& 0 & 1
\end{pmatrix}\,,\qquad 
P_{Rd}^*=
\begin{pmatrix}
e^{i(\phi_{a'_d}-\phi_{b_d}+\phi_{c_d})}  & 0&0  \\
0 & e^{i\phi_{b'_d}} &0 \\
0& 0 & e^{i\phi_{c_d}}
\end{pmatrix}\,.
\label{down-phase}
\end{align}
On the other hand, the up-type Yukawa matrix is also  given by the different two phase matrices and real Yukawa matrix:

\begin{align}
&  Y_u = P_{Lu}
\hat Y_u\, P_{Ru}^*\,, \qquad
\hat Y_u=
\begin{pmatrix}
0  &  a_u &0  \\
a'_u  & 0 & b_u \\
0& 0 & c_u 
\end{pmatrix}
,
\label{up-matrix}
\end{align}
where
\begin{align}
P_{Lu}=
\begin{pmatrix}
1  & 0&0  \\
0 & e^{i(\phi_{b_u}-\phi_{c_u})} &0 \\
0& 0 & 1
\end{pmatrix}\,,\qquad 
P_{Ru}^*=
\begin{pmatrix}
e^{i(\phi_{a'_u}-\phi_{b_u}+\phi_{c_u})}  & 0&0  \\
0 & e^{i\phi_{a_u}} &0 \\
0& 0 & e^{i\phi_{c_u}}
\end{pmatrix}\,.
\label{up-phase}
\end{align}
Then, the CKM matrix is given as:
\begin{align}
V_{\rm CKM}=
V_{Lu} P_{Lu}^* P_{Ld} V_{Ld}=V_{Lu} P_L V_{Ld}\,,
\label{CKM-model}
\end{align}
where
\begin{align}
P_{L}=
\begin{pmatrix}
1  & 0&0  \\
0 & e^{i\tau} &0 \\
0& 0 & 1
\end{pmatrix}\,.
\label{CKM-phase2}
\end{align}
We note that 
the phase $e^{i(\phi_{a_d}-\phi_{b'_d})}$ in
$P_{Ld}$ of Eq.~\eqref{down-phase} is absorbed by the u-quark field.
Thus, the CP phase is only $\tau$ in this model, which corresponds
to the CP phase in the CKM matrix.

The mass eigenvalues are given in terms of the real parameters
by solving eigenvalue equations of $\hat Y_d^T \hat Y_d$
and $\hat Y_u^T \hat Y_u$ as follows:
\begin{align}
\hat Y_d^T \hat Y_d=
\begin{pmatrix}
a_d^2  & 0& a_d\, b'_d  \\
0 &a'^2_d+b_d^2 & b_d\, c_d \\
a_d\, b'_d & b_d\, c_d& b'^2_d+c_d^2
\end{pmatrix}\,, \qquad
\hat Y_u^T \hat Y_u=
\begin{pmatrix}
a_u^2  & 0& 0 \\
0 &a'^2_u+b_u^2 & b_u\, c_u \\
0 & b_u\, c_u& c_u^2
\end{pmatrix}\,.
\label{MM}
\end{align}
We obtain quark masses as:
\begin{align}
&m_b^2\simeq v_d^2\, (b'^2_d+c_d^2)\,,\qquad 
m_s^2\simeq v_d^2\,\frac{b_d^2 \, b'^2_d}{ b'^2_d+c_d^2}\,,\qquad
m_d^2\simeq v_d^2\,\frac{a_d\, a'_d}{b_d\, b'_d}\,,\nonumber\\
&m_t^2\simeq v_u^2 \,c_u^2\,,\qquad\qquad\ \ \
m_c^2\simeq v_u^2\, a'^2_u\,,\qquad\qquad\ 
m_u^2\simeq v_u^2\,a_u^2\,,
\label{masses}
\end{align}
where $v_u$ and $v_d$ denote the  VEV of up-sector and down-sector Higgs bosons.
As seen in Eq.~\eqref{MM},
the Cabibbo angle $V_{us}$ and $V_{ub}$ arise from the down-type  Yukawa matrix while  $V_{us}$ comes from both the down-type
and up-type  Yukawa matrix;
that is 
\begin{align}
V_{us}\simeq\frac{a_d\, b'_d}{ b_d\, c_d}\,,
\qquad
V_{ub}\simeq\frac{a_d\, b'_d}{ b'^2_d+c_d^2}\,,\qquad
V_{cb}\simeq\frac{b_d\, c_d}{ b'^2_d+c_d^2}
-  e^{i\tau}\frac{b_u}{c_u}\,.
\end{align}
Thus, ten model parameters are fixed by inputting observed masses and CKM parameters at the electroweak scale \cite{Antusch:2013jca,ParticleDataGroup:2024cfk}.
An obtained sample set of our  numerical fitting is presented 
in Table \ref{tab:1}, which  are derived from the parameter set: 
 %%%%%%%%%%%%%%%%
 \begin{align}
 &\frac{a_d}{c_d}=0.0073\,,\qquad\quad\,\, \frac{a'_d}{c_d}=0.0044\,,\quad
 \frac{b_d}{c_d}=0.0382\,,\quad \frac{b'_d}{c_d}=0.815\,,\nonumber\\
 &\frac{a_u}{c_u}=7.75\times 10^{-6}\,,\quad \frac{a'_u}{c_u}=0.0035\,,\quad \frac{b_u}{c_u}=0.055\,,\qquad \tau=-46^\circ.
 \label{parameters-I}
 \end{align}

%%%%%%%%%%%%%%%%%%%%%%%
\begin{table}[h]
\centering
			\renewcommand{\arraystretch}{1.1}
\scalebox{0.84}{
            \begin{tabular}{|c|c|c|c|c|c|c|c|c|c|} \hline
				\rule[14pt]{0pt}{3pt}  
				& $\frac{m_s}{m_b}\hskip -1 mm\times\hskip -1 mm 10^2$ 
				& $\frac{m_d}{m_b}\hskip -1 mm\times\hskip -1 mm 10^4$& $\frac{m_c}{m_t}\hskip -1 mm\times\hskip -1 mm 10^3$&$\frac{m_u}{m_t}\hskip -1 mm\times\hskip -1 mm 10^6$&
				$|V_{us}|$ &$|V_{cb}|$ &$|V_{ub}|$&$|J_{\rm CP}|$& $\delta_{\rm CP}$
				\\
				\hline
				\rule[14pt]{0pt}{3pt}  
				fit &$1.95$ & $ 7.69$
				& $3.45$ & $7.74 $&
				$0.2229$ & $0.0418$ & $0.00357$ &
				$3.00\hskip -1 mm\times\hskip -1 mm 10^{-5}$&$68.6^\circ$
				\\ \hline
				\rule[14pt]{0pt}{3pt}
				Exp	 &$1.90$ & $9.64$ 
				& $3.65$& $ 7.50$ &
				$0.2250$ & $0.0418$ & $0.00373$ &$3.12\hskip -1 mm\times\hskip -1 mm 10^{-5}$&$65.72^\circ$\\
				$1\,\sigma$	&$\pm 0.12$ &$\pm 1.49$ & $\pm 0.15$& $^{+ 1.54}_{-3.05}$ &$ \pm 0.0007$ &
				$ ^{+ 0.00079}_{-0.00069}$ &
				$ {\pm 0.000090}$ &$^{+0.13}_{-0.12}\hskip -1 mm\times \hskip -1 mm 10^{-5}$&${\pm 1.50^\circ}$\\ \hline  
			\end{tabular}
        }
		\caption{A sample set  of the fit of the quark mass ratios, 
			CKM mixing angles,  $\delta_{\rm CP}$ and  $J_{\rm CP}$ 
	 at  EW scale. Experimental data are shown by the best fit 
      with  $1\, \sigma$ error-bar.}
		\label{tab:1}
\end{table}

%%%%%%%%%%%%%%%%%%%%%%
\subsection{Model $(a)$}
\label{Model-a}
Next, we consider 
the texture in Eq.~\eqref{(a)}, which  is explicitly written as:
\begin{align}
&  Y_q =
\begin{pmatrix}
 A_q & A'_q&0  \\
 0 & B_q &0 \\
0& B'_q& C_q
\end{pmatrix}_{LR}\,,
\label{Y(a)}
\end{align}
where the column exchange does not affect the masses and mixing angles.
The phase of each entry could be removed by the phase matrices
of the left- and right-handed quarks due to four zeros. This is easily understood in the similar discussion of subsection \ref{phase-structure}.

In order to obtain mass eigenvalues and the CP violating
observable, i.e., the Jarlskog invariant $J_{\mathrm{CP}}$ \cite{Jarlskog:1985ht},
we write down  $Y_q Y_q^\dagger$ as:
\begin{align}
&  Y_q Y_q^\dagger =
\begin{pmatrix}
 a_q^2+a_q^{'2} & a'_q b_q  & a'_q b'_q  \\
 a'_q b_q & b_q^2 & b_q b'_q \\
 a'_q b'_q & b_q b'_q &b^{'2}_q+c_q^2
\end{pmatrix}\,,
\label{YY(a)}
\end{align}
where $a_q$, $a'_q$, $b_q$ , $b'_q$ and $c_q$ are absolute values of 
 each entry of Eq.~\eqref{Y(a)}. 
The mass ratios are given as:
\begin{align}
\frac{m_d}{m_b}\sim   \frac{a_d}{c_d}\sim \lambda^4\,,
\qquad \frac{m_s}{m_b}\sim   \frac{b_d}{c_d}\sim \lambda^2\,,
\qquad 
\frac{m_u}{m_t}\sim   \frac{a_u}{c_u}\sim \lambda^8\,,
\qquad \frac{m_c}{m_t}\sim   \frac{b_u}{c_u}\sim \lambda^4\,,
\label{mass-ratio-a}
\end{align}
where $\lambda\simeq 0.2$ is put.
We can obtain approximately $J_{\mathrm{CP}}$ (see Eq.~\eqref{Jcp}
in Appendix \ref{app:Jcp} ) as
\begin{align}
J_{\mathrm{CP}} =
\frac{b'_d b'_u}{b_d^2 b_u^2 c_d^2 c_u^2}
\left[- b_db_u (a'^2_ub_d^2-a^2_d b^2_u)\sin\tau
+a'_da'_u\{b_d^2b_u^2\sin(2\tau-\sigma)-b_u^2b_d^2 \sin\tau\}
\right ]\,,
\end{align}
where phases are given in Eq.~\eqref{CKM-phase}.
Putting values of Eq.~\eqref{mass-ratio-a},
$J_{\mathrm{CP}}$ is smaller than $\lambda^8={\cal O}(10^{-6})$,
which is ruled  out by the observation $J_{\mathrm{CP}}\simeq  3\times 10^{-5}$. 
%%%%%%%%%%%%%%%%%%%%%%
%%%%%%%%%%%%%%%%%%%%%%%%%%%%%%
\subsection{Model $(b)$}
\label{Model-b}
The texture in Eq.~\eqref{(b)} is
  explicitly given by transposed matrix of Eq.~\eqref{Y(a)}.
Therefore, the mass ratios are the same ones
in Eq.~\eqref{mass-ratio-a}.
 We show   $Y_q Y_q^\dagger$ as:
\begin{align}
&  Y_q Y_q^\dagger =
\begin{pmatrix}
 a_q^2 & a_q a'_q  & 0  \\
 a'_q b_q &a_q^{'2}+ b_q^2 & b'_q c_q \\
 0 & b'_q c_q &c_q^2
\end{pmatrix}\,,
\label{YY(b)}
\end{align}
which has a different CKM structure from the model $(a)$ in
subsection \ref{Model-a}.
  The $J_{\mathrm{CP}}$ is  obtained approximately  as
\begin{align}
J_{\mathrm{CP}} \simeq
\frac{a_d a'_d a_u a'_u b'^2_d b'^2_u}{b_d^2 b_u^2 c_d^2 c_u^2}
\left[\sin(\tau+\sigma)-\sin(\tau-\sigma) \right ]
\,,
\end{align}
where phases are given in Eq.~\eqref{CKM-phase}.
Putting values of Eq.~\eqref{mass-ratio-a},
$J_{\mathrm{CP}}$ is also  smaller than $\lambda^8={\cal O}(10^{-6})$,
which is ruled  out by the observation $J_{\mathrm{CP}}\simeq  3\times 10^{-5}$. 
%%%%%%%%%%%%%%%%%%%%%%
%%%%%%%%%%%%%%%%%%%%%%%%%%%%%%
\subsection{Model $(c)$}

The texture of Eq.~\eqref{(c)} has the specific property that it has non-trivial CP phase leading to the weak CP, but such CP phase does not appear in its determinant. Therefore, this texture is available
to address the strong CP problem without axion.\footnote{The strong CP phase is written by $\bar \theta = \theta_{\rm QCD} + {\rm arg}({\rm det}Y_uY_d)$, where $\theta_{\rm QCD}$ is the coefficient of the QCD topological term. The texture $(c)$ and $(d)$ lead to ${\rm arg}({\rm det}Y_uY_d)=0$. We need $\theta_{\rm QCD}=0$ by some reason. For example, the moduli stabilization leading to $\theta_{\rm QCD}=0$ was discussed in Ref.~\cite{Higaki:2024pql}. }
Indeed, the quark mass matrix with this texture  has been discussed successfully in the framework of modular symmetry of flavors \cite{Feruglio:2023uof,Petcov:2024vph,Penedo:2024gtb}.

%%%%%%%%%%%%%%%%%%%%%%
%%%%%%%%%%%%%%%%%%%%%%%%%%%%%%
\subsection{Model $(d)$}
The down-type quark mass matrix of Eq.~\eqref{(d)}
was studied from the view point of phenomenology of the CKM matrix  \cite{Tanimoto:2016rqy}.
That is equivalent to the successful down-type quark matrix  $M_d^{(3)}$
with the diagonal up-type quark mass matrix in Ref.~\cite{Tanimoto:2016rqy}.

%%%%%%%%%%%%%%%%%%%%%%
%%%%%%%%%%%%%%%%%%%%%%%%%%%%%%%%%%%%%%%%%%%%%%%%%%%%%%%%%%%%%%%%%
\section{Conclusions}
\label{sec:con}
%%%%%%%%%%%%%%%%%%%%%%%%%%%%%%%%%%%%%%%%%%%%%%%%%%%%%%%%%%%%%%%%%

We have studied the coupling selection rules associated with non-group symmetries, i.e., $\mathbb{Z}_2$ gauging of $\mathbb{Z}_M$ symmetries. 
We have revealed the structure of Yukawa matrix obeying our selection rules for $M=3, 4$ and 5. 
It is found that the selection rules lead to various Yukawa textures as presented in sections \ref{sec:M=3} and 
 \ref{sec:M=4-5}, in which some of them cannot be realized by a conventional group symmetry. 
The typical one is the NNI texture, which has the simple CP phase
structure due to four zeros.
Indeed, the quark mass matrix of the NNI form works well phenomenologically
although twelve parameters fit the ten observables. 
We have also obtained the NNI-like texture, in which one entry vanishes
in the NNI form.
 Taking the NNI texture for the down-type quarks and
 the NNI-like one for up-type quarks, we present a viable model
 of the quark mass matrix with nine real parameters and one phase. 
 It is remarked that this model has not been discussed until now 
 because conventional symmetries could not derive those textures.

We have also discussed the  phenomenological implication of other successful textures. For instance, the texture of Eq.~\eqref{(c)} allows a solution of the strong CP problem without axion. 
Although we have studied the quark sector, it is interesting to apply our selection rules to the lepton sector.
Interesting textures were proposed in the lepton sector \cite{Frampton:2002yf,Fritzsch:2011qv}.
We would study derivations of textures in the lepton sector elsewhere. Furthermore, it would also be interesting to apply our proposed selection rules to the higher-dimensional operators in the framework of EFTs including the SMEFT, which is left for future work.

We comment on the possible ultraviolet origin of the selection rules discussed in this paper. 
Concerning the gauging of the automorphism $\mathbb{Z}_2$ of a single $\mathbb{Z}_M$ group, it appears in the context of 4D effective action of higher-dimensional Yang-Mills theory compactified on $T^2/\mathbb{Z}_2$ with magnetic fluxes \cite{Kobayashi:2024yqq} as mentioned in the Introduction, where the selection rule is the same with Eq. \eqref{eq:rule_before}. 
It would be interesting to extend this analysis to other higher-dimensional orbifolds including the outer automorphism $\mathbb{Z}_2 \times \mathbb{Z}_2^\prime$, which will lead to the selection rule in section \ref{sec:two_flavor}. We will leave it for future work.  

We have concentrated on deriving texture zeros by our selection rules.
We need hierarchical values of allowed entries to realize hierarchies of quark masses and mixing angles.
We may combine our selection rules, e.g. with the Froggatt-Nielsen mechanism \cite{Froggatt:1978nt}.
Also, a combination between our selection rules and modular flavor symmetries\footnote{For example, a realization of hierarchies without fine-tuning was studied in modular flavor symmetric models \cite{Feruglio:2021dte,Novichkov:2021evw,Ishiguro:2022pde,Petcov:2022fjf,Kikuchi:2023cap,Abe:2023ilq,Kikuchi:2023jap,Abe:2023qmr,Petcov:2023vws,Abe:2023dvr,deMedeirosVarzielas:2023crv,Kikuchi:2023dow,Kikuchi:2023fpl}.} would be interesting, because both may originate from string compactifications.
We would study it elsewhere.

\acknowledgments

We are very grateful to the referee for suggesting improvements. 
This work was supported by JSPS KAKENHI Grant Numbers JP23H04512 (H.O.) and JP23K03375 (T.K.).

\appendix

\section{Textures for $M=4$}
\label{app:M=4}

Here, we show the textures given by possible combinations of left- and right-handed quarks with flavor assignments i,$\cdots$, vii in Table~\ref{tab:M=4-1} when $M=4$.
Results are shown in Tables~\ref{tab:M=4-2}, \ref{tab:M=4-3} and \ref{tab:M=4-4}.
The combination of the flavor assignments ii and i for left-handed and right-handed quarks is omitted, because they correspond to the transposed matrices 
of the combination (i,ii).
Similar combinations are omitted.

\begin{table}[htbp]
\begin{center}
\begin{tabular} {|c|c|c|c|}  \hline 
Flavor (Left, Right) & Higgs$[g^0]$ & Higgs$[g^1]$ &     Higgs$[g^2]$  \\ \hline
 (i,i) & 
$\begin{pmatrix}
*&0&0\\
0&*&0\\
0&0&*
\end{pmatrix}$    & 
$\begin{pmatrix}
0&*&0\\
*&0&*\\
0&*&0
\end{pmatrix}$
 &           
$\begin{pmatrix}
0&0&*\\
0&*&0\\
*&0&0
\end{pmatrix}$
 \\
 \hline
(i,ii) &
$\begin{pmatrix}
*&*&0\\
0&0&*\\
0&0&0
\end{pmatrix}$ & 
$\begin{pmatrix}
0&0&*\\
*&*&0\\
0&0&*
\end{pmatrix}$ & 
$\begin{pmatrix}
0&0&0\\
0&0&*\\
*&*&0
\end{pmatrix}$ \\ \hline
(i,iii) &
$\begin{pmatrix}
*&*&0\\
0&0&0\\
0&0&*
\end{pmatrix}$ &
$\begin{pmatrix}
0&0&0\\
*&*&*\\
0&0&0
\end{pmatrix}$ &
$\begin{pmatrix}
0&0&*\\
0&0&0\\
*&*&0
\end{pmatrix}$ \\ \hline
(i,iv) & 
$\begin{pmatrix}
*&0&0\\
0&*&*\\
0&0&0
\end{pmatrix}$ &
$\begin{pmatrix}
0&*&*\\
*&0&0\\
0&*&*
\end{pmatrix}$ & 
$\begin{pmatrix}
0&0&0\\
0&*&*\\
*&0&0
\end{pmatrix}$ \\ \hline
(i,v) &
$\begin{pmatrix}
*&0&0\\
0&0&0\\
0&*&*
\end{pmatrix}$ &
$\begin{pmatrix}
0&0&0\\
*&*&*\\
0&0&0
\end{pmatrix}$ &
$\begin{pmatrix}
0&*&*\\
0&0&0\\
*&0&0
\end{pmatrix}$  \\ \hline
(i,vi) & 
$\begin{pmatrix}
0&0&0\\
*&*&0\\
0&0&*
\end{pmatrix}$ & 
$\begin{pmatrix}
*&*&0\\
0&0&*\\
*&*&0
\end{pmatrix}$ & 
$\begin{pmatrix}
0&0&*\\
*&*&0\\
0&0&0
\end{pmatrix}$ \\ \hline
(i,vii) & 
$\begin{pmatrix}
0&0&0\\
*&0&0\\
0&*&*
\end{pmatrix}$ &
$\begin{pmatrix}
*&0&0\\
0&*&*\\
*&0&0
\end{pmatrix}$ &
$\begin{pmatrix}
0&*&*\\
*&0&0\\
0&0&0
\end{pmatrix}$ \\ \hline
(ii,ii) & 
$\begin{pmatrix}
*&*&0\\
*&*&0\\
0&0&*
\end{pmatrix} $ & 
$\begin{pmatrix}
0&0&*\\
0&0&*\\
*&*&0
\end{pmatrix}$ &
$\begin{pmatrix}
0&0&0\\
0&0&0\\
0&0&*
\end{pmatrix}$ \\ \hline
(ii,iii) & 
$\begin{pmatrix}
*&*&0\\
*&*&0\\
0&0&0
\end{pmatrix}$ &
$\begin{pmatrix}
0&0&0\\
0&0&0\\
*&*&*
\end{pmatrix}$ &
$\begin{pmatrix}
0&0&*\\
0&0&*\\
0&0&0
\end{pmatrix}$ \\ \hline
(ii,iv) &
$\begin{pmatrix}
*&0&0\\
*&0&0\\
0&*&*
\end{pmatrix}$ & 
$\begin{pmatrix}
0&*&*\\
0&*&*\\
*&0&0
\end{pmatrix}$ & 
$\begin{pmatrix}
0&0&0\\
0&0&0\\
0&*&*
\end{pmatrix}$ \\ \hline
(ii,v) & 
$\begin{pmatrix}
*&0&0\\
*&0&0\\
0&0&0
\end{pmatrix}$ &
$\begin{pmatrix}
0&0&0\\
0&0&0\\
*&*&*
\end{pmatrix}$ &
$\begin{pmatrix}
0&*&*\\
0&*&*\\
0&0&0
\end{pmatrix}$ \\ \hline
(ii, vi) &
$\begin{pmatrix}
0&0&0\\
0&0&0\\
*&*&0
\end{pmatrix}$ &
$\begin{pmatrix}
*&*&0\\
*&*&0\\
0&0&*
\end{pmatrix}$ &
$\begin{pmatrix}
0&0&*\\
0&0&*\\
*&*&0
\end{pmatrix}$ \\ \hline
\end{tabular}
\caption{Yukawa textures for $M=4$.}
\label{tab:M=4-2}
\end{center}
\end{table}

\begin{table}[htbp]
\begin{center}
\begin{tabular} {|c|c|c|c|}  \hline 
Flavor (Left, Right) & Higgs$[g^0]$ & Higgs$[g^1]$ &     Higgs$[g^2]$  \\ \hline
(ii, vii) & 
$\begin{pmatrix}
0&0&0\\
0&0&0\\
*&0&0
\end{pmatrix}$ & 
$\begin{pmatrix}
*&0&0\\
*&0&0\\
0&*&*
\end{pmatrix}$ &
$\begin{pmatrix}
0&*&*\\
0&*&*\\
*&0&0
\end{pmatrix}$ \\ \hline
(iii,iii) &
$\begin{pmatrix}
*&*&0\\
*&*&0\\
0&0&*
\end{pmatrix}$ & 
$\begin{pmatrix}
0&0&0\\
0&0&0\\
0&0&0
\end{pmatrix}$ &
$\begin{pmatrix}
0&0&*\\
0&0&*\\
*&*&0
\end{pmatrix}$ \\ \hline
(iii,iv) &
$\begin{pmatrix}
*&0&0\\
*&0&0\\
0&0&0
\end{pmatrix}$ &
$\begin{pmatrix}
0&*&*\\
0&*&*\\
0&*&*
\end{pmatrix}$ &
$\begin{pmatrix}
0&0&0\\
0&0&0\\
*&0&0
\end{pmatrix}$ \\ \hline
(iii,v) &
$\begin{pmatrix}
*&0&0\\
*&0&0\\
0&*&*
\end{pmatrix}$ &
$\begin{pmatrix}
0&0&0\\
0&0&0\\
0&0&0
\end{pmatrix}$ &
$\begin{pmatrix}
0&*&*\\
0&*&*\\
*&0&0
\end{pmatrix}$ \\ \hline 
(iii,vi) & 
$\begin{pmatrix}
0&0&0\\
0&0&0\\
0&0&*
\end{pmatrix}$ & 
$\begin{pmatrix}
*&*&0\\
*&*&0\\
*&*&0
\end{pmatrix}$ & 
$\begin{pmatrix}
0&0&*\\
0&0&*\\
0&0&0
\end{pmatrix}$ \\ \hline
(iii,vii) & 
$\begin{pmatrix}
0&0&0\\
0&0&0\\
0&*&*
\end{pmatrix}$ &
$\begin{pmatrix}
*&0&0\\
*&0&0\\
*&0&0
\end{pmatrix}$ &
$\begin{pmatrix}
0&*&*\\
0&*&*\\
0&0&0
\end{pmatrix}$ \\ \hline
(iv,iv) &
$\begin{pmatrix}
*&0&0\\
0&*&*\\
0&*&*
\end{pmatrix}$ &
$\begin{pmatrix}
0&*&*\\
*&0&0\\
*&0&0
\end{pmatrix}$ &
$\begin{pmatrix}
0&0&0\\
0&*&*\\
0&*&*
\end{pmatrix}$ \\ \hline
(iv,v) & 
$\begin{pmatrix}
*&0&0\\
0&0&0\\
0&0&0
\end{pmatrix}$ &
$\begin{pmatrix}
0&0&0\\
*&*&*\\
*&*&*
\end{pmatrix}$ &
$\begin{pmatrix}
0&*&*\\
0&0&0\\
0&0&0
\end{pmatrix}$ \\ \hline
(iv,vi) &
$\begin{pmatrix}
0&0&0\\
*&*&0\\
*&*&0
\end{pmatrix}$ &
$\begin{pmatrix}
*&*&0\\
0&0&*\\
0&0&*
\end{pmatrix}$ &
$\begin{pmatrix}
0&0&*\\
*&*&0\\
*&*&0
\end{pmatrix}$ \\ \hline
(iv,vii) & 
$\begin{pmatrix}
0&0&0\\
*&0&0\\
*&0&0
\end{pmatrix}$ &
$\begin{pmatrix}
*&0&0\\
0&*&*\\
0&*&*
\end{pmatrix}$ &
$\begin{pmatrix}
0&*&*\\
*&0&0\\
*&0&0
\end{pmatrix}$ \\ \hline 
(v,v) &
$\begin{pmatrix}
*&0&0\\
0&*&*\\
0&*&*
\end{pmatrix}$ &
$\begin{pmatrix}
0&0&0\\
0&0&0\\
0&0&0
\end{pmatrix}$ &
$\begin{pmatrix}
0&*&*\\
*&0&0\\
*&0&0
\end{pmatrix}$ \\ \hline 
(v,vi) & 
$\begin{pmatrix}
0&0&0\\
0&0&*\\
0&0&*
\end{pmatrix}$ &
$\begin{pmatrix}
*&*&0\\
*&*&0\\
*&*&0
\end{pmatrix}$ &
$\begin{pmatrix}
0&0&*\\
0&0&0\\
0&0&0
\end{pmatrix}$ \\ \hline
\end{tabular}
\caption{Yukawa textures for $M=4$.}
\label{tab:M=4-3}
\end{center}
\end{table}

\clearpage

\begin{table}[htbp]
\begin{center}
\begin{tabular} {|c|c|c|c|}  \hline 
Flavor (Left, Right) & Higgs$[g^0]$ & Higgs$[g^1]$ &     Higgs$[g^2]$  \\ \hline
(v,vii) &
$\begin{pmatrix}
0&0&0\\
0&*&*\\
0&*&*
\end{pmatrix}$ &
$\begin{pmatrix}
*&0&0\\
*&0&0\\
*&0&0
\end{pmatrix}$ &
$\begin{pmatrix}
0&*&*\\
0&0&0\\
0&0&0
\end{pmatrix}$ \\ \hline 
(vi,vi) & 
$\begin{pmatrix}
*&*&0\\
*&*&0\\
0&0&*
\end{pmatrix}$ &
$\begin{pmatrix}
0&0&*\\
0&0&*\\
*&*&0
\end{pmatrix}$ &
$\begin{pmatrix}
*&*&0\\
*&*&0\\
0&0&0
\end{pmatrix}$ \\ \hline
(vi,vii) &
$\begin{pmatrix}
*&0&0\\
*&0&0\\
0&*&*
\end{pmatrix}$ &
$\begin{pmatrix}
0&*&*\\
0&*&*\\
*&0&0
\end{pmatrix}$ &
$\begin{pmatrix}
*&0&0\\
*&0&0\\
0&0&0
\end{pmatrix}$ \\ \hline
(vii,vii) & 
$\begin{pmatrix}
*&0&0\\
0&*&*\\
0&*&*
\end{pmatrix}$ &
$\begin{pmatrix}
0&*&*\\
*&0&0\\
*&0&0
\end{pmatrix}$ &
$\begin{pmatrix}
*&0&0\\
0&0&0\\
0&0&0
\end{pmatrix}$ \\ \hline
\end{tabular}
\caption{Yukawa textures for $M=4$.}
\label{tab:M=4-4}
\end{center}
\end{table}

\section{Textures for $M=5$}
\label{app:M=5}

Here, we show the textures given by possible combinations of left- and right-handed quarks with flavor assignments i,$\cdots$, vii in Table~\ref{tab:M=5-1} when $M=5$.
Results are shown in Tables~\ref{tab:M=5-2}, \ref{tab:M=5-3} and \ref{tab:M=5-4}.
The combination of the flavor assignments ii and i for left-handed and right-handed quarks is omitted, because they correspond to the transposed matrices 
of the combination (i,ii).
Similar combinations are omitted.

\begin{table}[htbp]
\begin{center}
\begin{tabular} {|c|c|c|c|}  \hline 
Flavor (Left, Right) & Higgs$[g^0]$ & Higgs$[g^1]$ &     Higgs$[g^2]$  \\ \hline
 (i,i) & 
$\begin{pmatrix}
*&0&0\\
0&*&0\\
0&0&*
\end{pmatrix}$    & 
$\begin{pmatrix}
0&*&0\\
*&0&*\\
0&*&*
\end{pmatrix}$
 &           
$\begin{pmatrix}
0&0&*\\
0&*&*\\
*&*&0
\end{pmatrix}$
 \\
 \hline
(i,ii) &
$\begin{pmatrix}
*&*&0\\
0&0&*\\
0&0&0
\end{pmatrix}$ & 
$\begin{pmatrix}
0&0&*\\
*&*&0\\
0&0&*
\end{pmatrix}$ & 
$\begin{pmatrix}
0&0&0\\
0&0&*\\
*&*&*
\end{pmatrix}$ \\ \hline
(i,iii) &
$\begin{pmatrix}
*&*&0\\
0&0&0\\
0&0&*
\end{pmatrix}$ &
$\begin{pmatrix}
0&0&0\\
*&*&*\\
0&0&*
\end{pmatrix}$ &
$\begin{pmatrix}
0&0&*\\
0&0&*\\
*&*&0
\end{pmatrix}$ \\ \hline
(i,iv) & 
$\begin{pmatrix}
*&0&0\\
0&*&*\\
0&0&0
\end{pmatrix}$ &
$\begin{pmatrix}
0&*&*\\
*&0&0\\
0&*&*
\end{pmatrix}$ & 
$\begin{pmatrix}
0&0&0\\
0&*&*\\
*&*&*
\end{pmatrix}$ \\ \hline
(i,v) &
$\begin{pmatrix}
*&0&0\\
0&0&0\\
0&*&*
\end{pmatrix}$ &
$\begin{pmatrix}
0&0&0\\
*&*&*\\
0&*&*
\end{pmatrix}$ &
$\begin{pmatrix}
0&*&*\\
0&*&*\\
*&0&0
\end{pmatrix}$  \\ \hline
(i,vi) & 
$\begin{pmatrix}
0&0&0\\
*&*&0\\
0&0&*
\end{pmatrix}$ & 
$\begin{pmatrix}
*&*&0\\
0&0&*\\
*&*&*
\end{pmatrix}$ & 
$\begin{pmatrix}
0&0&*\\
*&*&*\\
*&*&0
\end{pmatrix}$ \\ \hline
(i,vii) & 
$\begin{pmatrix}
0&0&0\\
*&0&0\\
0&*&*
\end{pmatrix}$ &
$\begin{pmatrix}
*&0&0\\
0&*&*\\
*&*&*
\end{pmatrix}$ &
$\begin{pmatrix}
0&*&*\\
*&*&*\\
*&0&0
\end{pmatrix}$ \\ \hline
(ii,ii) & 
$\begin{pmatrix}
*&*&0\\
*&*&0\\
0&0&*
\end{pmatrix} $ & 
$\begin{pmatrix}
0&0&*\\
0&0&*\\
*&*&0
\end{pmatrix}$ &
$\begin{pmatrix}
0&0&0\\
0&0&0\\
0&0&*
\end{pmatrix}$ \\ \hline
(ii,iii) & 
$\begin{pmatrix}
*&*&0\\
*&*&0\\
0&0&0
\end{pmatrix}$ &
$\begin{pmatrix}
0&0&0\\
0&0&0\\
*&*&*
\end{pmatrix}$ &
$\begin{pmatrix}
0&0&*\\
0&0&*\\
0&0&*
\end{pmatrix}$ \\ \hline
(ii,iv) &
$\begin{pmatrix}
*&0&0\\
*&0&0\\
0&*&*
\end{pmatrix}$ & 
$\begin{pmatrix}
0&*&*\\
0&*&*\\
*&0&0
\end{pmatrix}$ & 
$\begin{pmatrix}
0&0&0\\
0&0&0\\
0&*&*
\end{pmatrix}$ \\ \hline
(ii,v) & 
$\begin{pmatrix}
*&0&0\\
*&0&0\\
0&0&0
\end{pmatrix}$ &
$\begin{pmatrix}
0&0&0\\
0&0&0\\
*&*&*
\end{pmatrix}$ &
$\begin{pmatrix}
0&*&*\\
0&*&*\\
0&*&*
\end{pmatrix}$ \\ \hline
(ii, vi) &
$\begin{pmatrix}
0&0&0\\
0&0&0\\
*&*&0
\end{pmatrix}$ &
$\begin{pmatrix}
*&*&0\\
*&*&0\\
0&0&*
\end{pmatrix}$ &
$\begin{pmatrix}
0&0&*\\
0&0&*\\
*&*&*
\end{pmatrix}$ \\ \hline
\end{tabular}
\caption{Yukawa textures for $M=5$.}
\label{tab:M=5-2}
\end{center}
\end{table}

\begin{table}[htbp]
\begin{center}
\begin{tabular} {|c|c|c|c|}  \hline 
Flavor (Left, Right) & Higgs$[g^0]$ & Higgs$[g^1]$ &     Higgs$[g^2]$  \\ \hline
(ii, vii) & 
$\begin{pmatrix}
0&0&0\\
0&0&0\\
*&0&0
\end{pmatrix}$ & 
$\begin{pmatrix}
*&0&0\\
*&0&0\\
0&*&*
\end{pmatrix}$ &
$\begin{pmatrix}
0&*&*\\
0&*&*\\
*&*&*
\end{pmatrix}$ \\ \hline
(iii,iii) &
$\begin{pmatrix}
*&*&0\\
*&*&0\\
0&0&*
\end{pmatrix}$ & 
$\begin{pmatrix}
0&0&0\\
0&0&0\\
0&0&*
\end{pmatrix}$ &
$\begin{pmatrix}
0&0&*\\
0&0&*\\
*&*&0
\end{pmatrix}$ \\ \hline
(iii,iv) &
$\begin{pmatrix}
*&0&0\\
*&0&0\\
0&0&0
\end{pmatrix}$ &
$\begin{pmatrix}
0&*&*\\
0&*&*\\
0&*&*
\end{pmatrix}$ &
$\begin{pmatrix}
0&0&0\\
0&0&0\\
*&*&*
\end{pmatrix}$ \\ \hline
(iii,v) &
$\begin{pmatrix}
*&0&0\\
*&0&0\\
0&*&*
\end{pmatrix}$ &
$\begin{pmatrix}
0&0&0\\
0&0&0\\
0&*&*
\end{pmatrix}$ &
$\begin{pmatrix}
0&*&*\\
0&*&*\\
*&0&0
\end{pmatrix}$ \\ \hline 
(iii,vi) & 
$\begin{pmatrix}
0&0&0\\
0&0&0\\
0&0&*
\end{pmatrix}$ & 
$\begin{pmatrix}
*&*&0\\
*&*&0\\
*&*&*
\end{pmatrix}$ & 
$\begin{pmatrix}
0&0&*\\
0&0&*\\
*&*&0
\end{pmatrix}$ \\ \hline
(iii,vii) & 
$\begin{pmatrix}
0&0&0\\
0&0&0\\
0&*&*
\end{pmatrix}$ &
$\begin{pmatrix}
*&0&0\\
*&0&0\\
*&*&*
\end{pmatrix}$ &
$\begin{pmatrix}
0&*&*\\
0&*&*\\
*&0&0
\end{pmatrix}$ \\ \hline
(iv,iv) &
$\begin{pmatrix}
*&0&0\\
0&*&*\\
0&*&*
\end{pmatrix}$ &
$\begin{pmatrix}
0&*&*\\
*&0&0\\
*&0&0
\end{pmatrix}$ &
$\begin{pmatrix}
0&0&0\\
0&*&*\\
0&*&*
\end{pmatrix}$ \\ \hline
(iv,v) & 
$\begin{pmatrix}
*&0&0\\
0&0&0\\
0&0&0
\end{pmatrix}$ &
$\begin{pmatrix}
0&0&0\\
*&*&*\\
*&*&*
\end{pmatrix}$ &
$\begin{pmatrix}
0&*&*\\
0&*&*\\
0&*&*
\end{pmatrix}$ \\ \hline
(iv,vi) &
$\begin{pmatrix}
0&0&0\\
*&*&0\\
*&*&0
\end{pmatrix}$ &
$\begin{pmatrix}
*&*&0\\
0&0&*\\
0&0&*
\end{pmatrix}$ &
$\begin{pmatrix}
0&0&*\\
*&*&*\\
*&*&*
\end{pmatrix}$ \\ \hline
(iv,vii) & 
$\begin{pmatrix}
0&0&0\\
*&0&0\\
*&0&0
\end{pmatrix}$ &
$\begin{pmatrix}
*&0&0\\
0&*&*\\
0&*&*
\end{pmatrix}$ &
$\begin{pmatrix}
0&*&*\\
*&*&*\\
*&*&*
\end{pmatrix}$ \\ \hline 
(v,v) &
$\begin{pmatrix}
*&0&0\\
0&*&*\\
0&*&*
\end{pmatrix}$ &
$\begin{pmatrix}
0&0&0\\
0&*&*\\
0&*&*
\end{pmatrix}$ &
$\begin{pmatrix}
0&*&*\\
*&0&0\\
*&0&0
\end{pmatrix}$ \\ \hline 
(v,vi) & 
$\begin{pmatrix}
0&0&0\\
0&0&*\\
0&0&*
\end{pmatrix}$ &
$\begin{pmatrix}
*&*&0\\
*&*&*\\
*&*&*
\end{pmatrix}$ &
$\begin{pmatrix}
0&0&*\\
*&*&0\\
*&*&0
\end{pmatrix}$ \\ \hline
\end{tabular}
\caption{Yukawa textures for $M=5$.}
\label{tab:M=5-3}
\end{center}
\end{table}

\clearpage

\begin{table}[htbp]
\begin{center}
\begin{tabular} {|c|c|c|c|}  \hline 
Flavor (Left, Right) & Higgs$[g^0]$ & Higgs$[g^1]$ &     Higgs$[g^2]$  \\ \hline
(v,vii) &
$\begin{pmatrix}
0&0&0\\
0&*&*\\
0&*&*
\end{pmatrix}$ &
$\begin{pmatrix}
*&0&0\\
*&*&*\\
*&*&*
\end{pmatrix}$ &
$\begin{pmatrix}
0&*&*\\
*&0&0\\
*&0&0
\end{pmatrix}$ \\ \hline 
(vi,vi) & 
$\begin{pmatrix}
*&*&0\\
*&*&0\\
0&0&*
\end{pmatrix}$ &
$\begin{pmatrix}
0&0&*\\
0&0&*\\
*&*&*
\end{pmatrix}$ &
$\begin{pmatrix}
*&*&*\\
*&*&*\\
*&*&0
\end{pmatrix}$ \\ \hline
(vi,vii) &
$\begin{pmatrix}
*&0&0\\
*&0&0\\
0&*&*
\end{pmatrix}$ &
$\begin{pmatrix}
0&*&*\\
0&*&*\\
*&*&*
\end{pmatrix}$ &
$\begin{pmatrix}
*&*&*\\
*&*&*\\
*&0&0
\end{pmatrix}$ \\ \hline
(vii,vii) & 
$\begin{pmatrix}
*&0&0\\
0&*&*\\
0&*&*
\end{pmatrix}$ &
$\begin{pmatrix}
0&*&*\\
*&*&*\\
*&*&*
\end{pmatrix}$ &
$\begin{pmatrix}
*&*&*\\
*&0&0\\
*&0&0
\end{pmatrix}$ \\ \hline
\end{tabular}
\caption{Yukawa textures for $M=5$.}
\label{tab:M=5-4}
\end{center}
\end{table}

\section{Jarlskog invariant $J_{\mathrm{CP}}$}
\label{app:Jcp}
There is the CP violating observable, the Jarlskog invariant $J_{\mathrm{CP}}$ \cite{Jarlskog:1985ht},
which is derived from the following relation:
\begin{align}
&iC\equiv [M_u M_u^\dagger, M_d M_d^\dagger]  \ , \nonumber \\
& \det C= -2 J_{\mathrm{CP}}\, (m_t^2-m_c^2) (m_c^2-m_u^2) (m_u^2-m_t^2) (m_b^2-m_s^2) (m_s^2-m_d^2) (m_d^2-m_b^2) \ .
\label{Jcp}
\end{align}

\bibliography{references}{}
\bibliographystyle{JHEP}

\end{document}